\documentclass[reprint,aps,epsfig,superscriptaddress,twocolumn]{revtex4}

\usepackage{bm}
\usepackage{times}
\usepackage[dvips]{graphicx}
\usepackage{mathrsfs}
\usepackage[intlimits]{amsmath}
\usepackage[colorlinks, citecolor=red]{hyperref}
\usepackage[switch]{lineno}
\usepackage{algorithm}
\usepackage{algpseudocode}
\usepackage{setspace}
\usepackage{xcolor}
\usepackage{color}
\usepackage[defaultcolr=black]{changes}
\usepackage{ulem}
\usepackage{tabularx}

\newcommand{\ket}[1]{\vert #1 \rangle}
\definechangesauthor[name={Author1}, color=red]{A2}

\begin{document}

\title{Realization of a programmable multipurpose photonic quantum memory with over-thousand qubit manipulations}

\author{Sheng Zhang}

\affiliation{Center for Quantum Information, Institute for Interdisciplinary Information Sciences, Tsinghua University, Beijing 100084, PR China}

\author{Jixuan Shi}

\affiliation{Center for Quantum Information, Institute for Interdisciplinary Information Sciences, Tsinghua University, Beijing 100084, PR China}

\author{Zhaibin Cui}

\affiliation{Center for Quantum Information, Institute for Interdisciplinary Information Sciences, Tsinghua University, Beijing 100084, PR China}

\author{Ye Wang}

\affiliation{Center for Quantum Information, Institute for Interdisciplinary Information Sciences, Tsinghua University, Beijing 100084, PR China}

\author{Yukai Wu}

\affiliation{Center for Quantum Information, Institute for Interdisciplinary Information Sciences, Tsinghua University, Beijing 100084, PR China}
\affiliation{Hefei National Laboratory, Hefei 230088, PR China}

\author{Luming Duan}
\email{lmduan@tsinghua.edu.cn}
\affiliation{Center for Quantum Information, Institute for Interdisciplinary Information Sciences, Tsinghua University, Beijing 100084, PR China}
\affiliation{Hefei National Laboratory, Hefei 230088, PR China}

\author{Yunfei Pu}
\email{puyf@tsinghua.edu.cn}
\affiliation{Center for Quantum Information, Institute for Interdisciplinary Information Sciences, Tsinghua University, Beijing 100084, PR China}
\affiliation{Hefei National Laboratory, Hefei 230088, PR China}

\begin{abstract}
  Quantum networks can enable various applications such as distributed quantum computing, long-distance quantum communication, and network-based quantum sensing with unprecedented performances. One of the most important building blocks for a quantum network is a photonic quantum memory which serves as the interface between the communication channel and the local functional unit. A programmable quantum memory which can process a large stream of flying qubits and fulfill the requirements of multiple core functions in a quantum network is still to-be-realized. Here we report a high-performance quantum memory which can simultaneously store $72$ optical qubits carried by $144$ spatially separated atomic ensembles  and support up to a thousand consecutive write or read operations in a random access way, two orders of magnitude larger than the previous record. Due to the built-in programmability, this quantum memory can be adapted on-demand for several functions. As example applications, we realize quantum queue, stack, and buffer which closely resemble the counterpart devices for classical information processing. We further demonstrate the storage and reshuffle of $4$ entangled pairs of photonic pulses with probabilistic arrival time and arbitrary release order via the memory, which is an essential requirement for the realization of quantum repeaters and efficient routing in quantum networks. Realization of this multi-purpose programmable quantum memory thus constitutes a key enabling building block for future large-scale fully-functional quantum networks.
\end{abstract}

\maketitle

\section{Introduction}

 \noindent Quantum memory is a basic ingredient for quantum information technology and plays a key role in quantum computing {\cite{ion, sc}}, communication \cite{BDCZ, DLCZ, oxford_qkd, weifurner_qkd} and networking~\cite{kimble, weiner, distributed, oxford_qkd, weifurner_qkd, hanson_teleportation, ye_and_lukin, repeater_telescope, rmp_gisin, rmp_polzik}. In a quantum network based on the quantum repeater architecture, photonic quantum memory serves as the local quantum nodes to connect the communication channels and thus provides the key building block, as shown in Fig.~1(a). For future applications in large-scale quantum computing and networking, it is desirable that the quantum memory is versatile to support application-oriented operations and fits for tasks in different scenarios \cite{quantum_os,linklayer, kimble, weiner}.

Substantial advances have been made for the realization of various quantum memories. Photonic quantum memory with long coherence time has been achieved based on atomic or spin ensembles \cite{1min, bao0.1s, zhongmanjin, udisk, hot_atom}, single atoms or ions \cite{rempe0.1s, oxford_sr_and_ca}, and single vacancy centers \cite{lukin_siv}. To scale up the capacity of quantum memory and boost the performance of a quantum repeater~\cite{07prl, rmp_gisin, simon, 2021nature}, multiplexed quantum memory has been proposed and implemented via spatial modes \cite{225, lan}, temporal modes \cite{1250, 1650}, orbital angular momentum modes \cite{dingdongshen, laurent}, wavevector directions \cite{shanxi, parniak}, and other dimensions \cite{wurst, 1650, kwiat}. In the multiplexed quantum repeater protocol, a quantum memory capable of reading out the stored qubits on-demand is an indispensable ingredient to achieve the full-scale acceleration in entangling rate \cite{07prl, rmp_gisin, simon}. Random access quantum memories (RAQMs) \cite{jiangnan,rempe_ram, wurst, wolters} developed recently are perfectly suitable for this task. To achieve a reasonable entangling rate over a metropolitan size of $100\,$km via the quantum repeater protocol, one has to combine long coherence time, large memory capacity, and high-fidelity operations altogether. Although there are demonstrations for each individual technique, how to make them compatible with each other and integrate all these elements into a single quantum memory setup remains a challenging goal in experiment.

\begin{figure*}
  \centering
  \includegraphics[width=17.6cm]{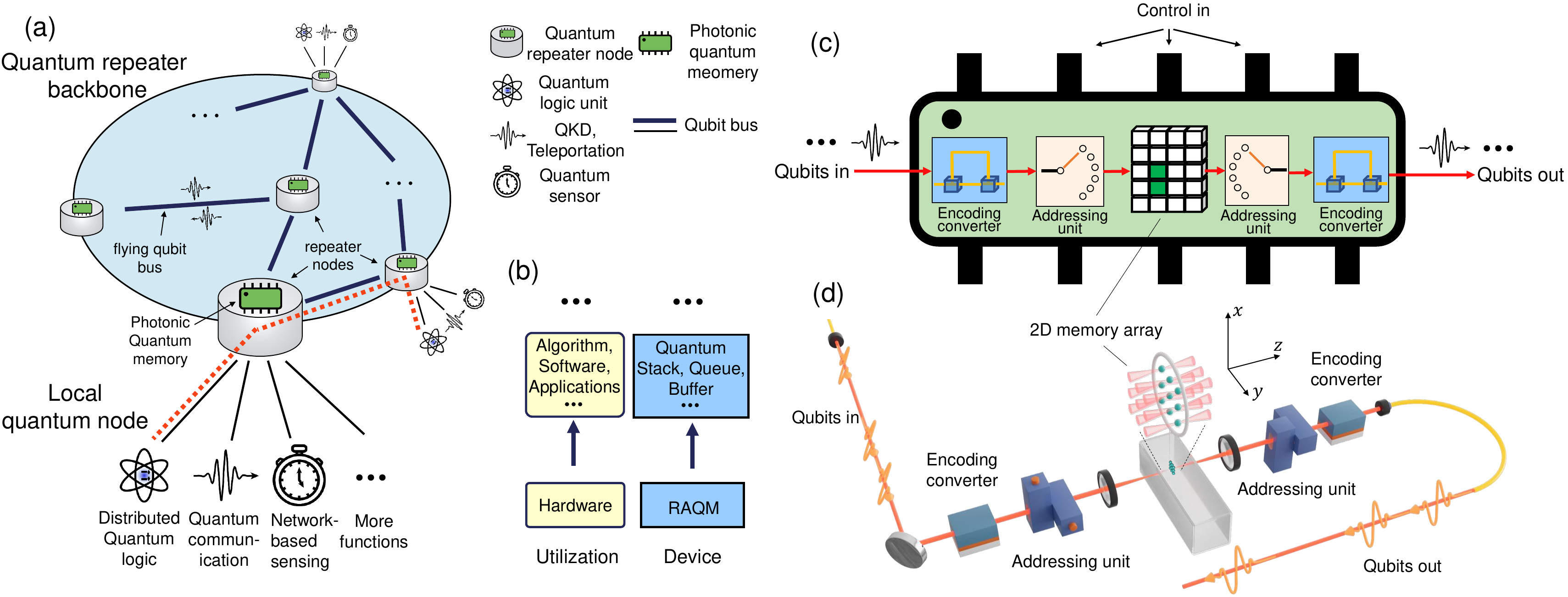}\\
  \caption{Multi-purpose photonic quantum memory and experimental setup.
  (a) Quantum network architecture with a communication backbone based on a quantum repeater, and many local quantum nodes. The red dashed lines demonstrate the connection scheme for distributed quantum computing based on this architecture. (b) Different types of quantum data containers and their utilizations. (c)-(d) Schematic of the multipurpose quantum memory. The system consists of two encoding converters, two addressing units, and a two-dimensional memory array based on a $^{87}$Rb atomic cloud. Both the input and output qubits are encoded in polarization.
  }
\end{figure*}

In this work, we realize a photonic quantum memory with a capacity of $72$ qubit cells (size of a state-of-the-art quantum processor), coherence time above $0.5\,$ms for almost all cells (corresponding to the transmission time in a $100\,$km fiber), and $1000$ consecutive operations in a random access way (sufficient to generate any of the $72!\approx10^{104}$ output sequences for a $72$-qubit input), two orders of magnitude larger than the previous record \cite{rempe_ram, wolters}. We further demonstrate quantum queue, stack, and buffer, which are the quantum counterparts of the corresponding widely-used classical memory devices. Finally, as an example of applications, we demonstrate the storage and reshuffle of four entangled photonic pairs with probabilistic arrival time and arbitrary release order for the first time, which can find applications in several scenarios of a quantum network including entanglement generation, connection, and routing. The photonic quantum memory implemented here, which is scalable, long-lived, fully-controllable, and combines unprecedented versatility, provides key enabling ingredients for the implementation of quantum repeaters and large-scale quantum networks.

\section{Results}

\subsection{Experimental setup}

The multipurpose quantum memory consists of a two-dimensional quantum memory array, two addressing units and two qubit encoding converters, as illustrated in Fig.~1(c) and 1(d). The 2D quantum memory array is based on a macroscopic $^{87}$Rb atomic cloud divided to 12$\times$12=144 pieces spatially, and each of the $144$ microscopic ensembles represents a quantum memory. Here different memory cells are different parts of a single atomic cloud, and these $144$ cells are not loaded into optical trap arrays, which is different from the tweezer array experiments~\cite{tweezer1,tweezer2}. $3$ pairs of crossed acoustic optical deflectors (AODs) act as addressing units and are exploited to direct the input/output photon and control beam (not shown in Fig.~1) to different memory cells. Here we use a pair of micro-ensembles to store a qubit, and the detailed pairing strategy of the 72 qubit cells out of the 144 micro-ensembles is illustrated in Fig.~2(d). The detailed implementation of the experimental setup is described in Appendix A.

In the storage process, the photonic qubit is coherently converted to a spin wave excitation by electromagnetic induced transparency (EIT) \cite{kuzmich_eit}. Here, we use a pair of clock states $|5S_{1/2},F=1, m_F=0\rangle$ and $|5S_{1/2},F=2, m_F=0\rangle$ to cancel the decoherence induced by magnetic field noise, and a collinear configuration for the optical beams to suppress the decoherence caused by atomic random motion respectively, which elongates the coherence time of each memory cell to over 500$\mu$s (Fig.~2(a)). Together with a much faster access time of 1$\mu$s through upgraded optical design (see Appendix D), we can significantly improve the maximally allowed operations limited by the ratio of memory coherence time to the access time, from previously state-of-art 12 to $1000$. We further measure the storage efficiency  (including both write and read in  atoms, the detailed efficiencies of all components are listed in Table I) in each of the $144$ micro-ensembles, as illustrated in Fig.~2(b). We also measure the storage fidelity in $6$ qubit cells located in different regions, with $4$ complementary states $|H\rangle$, $|V\rangle$, $|+\rangle=\frac{|H\rangle+|V\rangle}{\sqrt{2}}$, and $|L\rangle=\frac{|H\rangle+i|V\rangle}{\sqrt{2}}$ as input~\cite{zhushiliang} and different storage time from 100$\mu$s to 500$\mu$s, as shown in Fig. 2(e) and 2(f). Note that the input polarization can be arbitrary, as shown in Fig. 2(c). The average fidelity over 4 different input states is above $95\%$. Furthermore, the crosstalk is measured to be about $1\%$ by a round of operations on the 6 neighboring micro-ensembles surrounding the target qubit (Fig.~2(g)). Finally, repeated writing and reading on the same qubit cell is found to have negligible influence to the performance.


\begin{figure*}
  \centering
  \includegraphics[width=17.4cm]{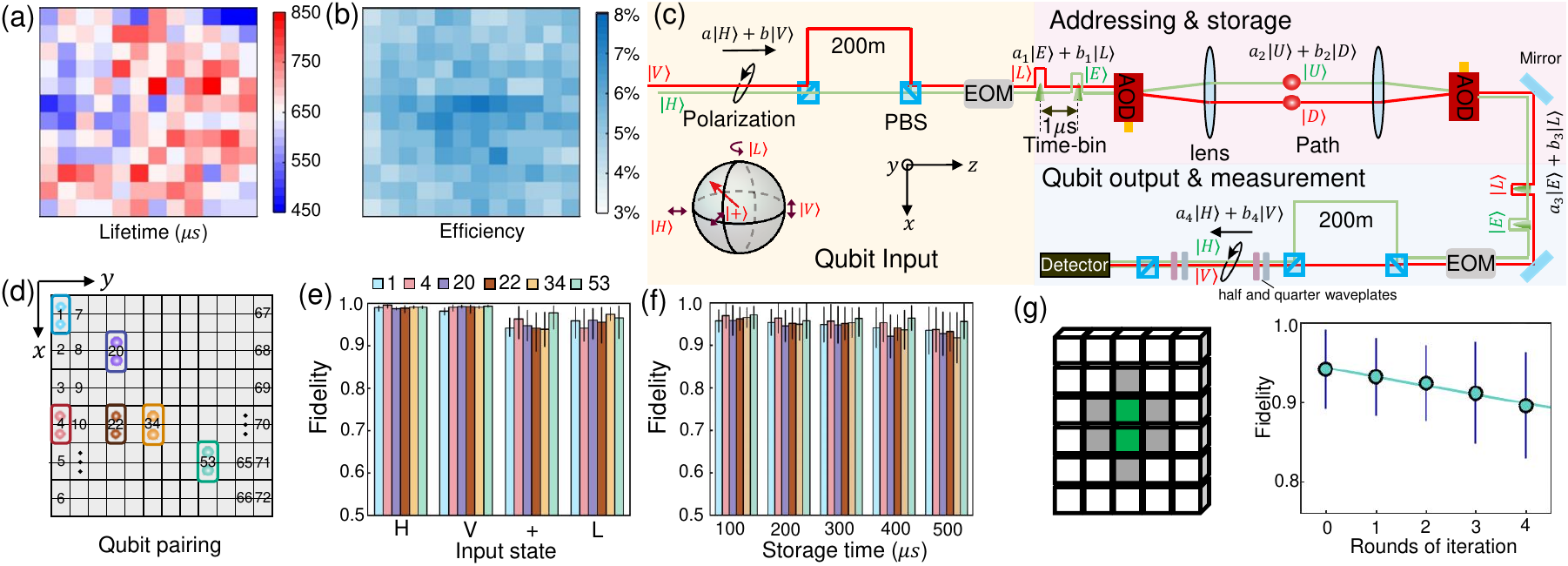}\\
  \caption{Performance of individual memory units.
  (a),(b), Coherence time and storage efficiency of the 144 micro-ensembles. (c), The storage and retrieval of an arbitrary polarization qubit. We first convert a polarization qubit $a|H\rangle+b|V\rangle=cos(\theta)|H\rangle+e^{i\phi}sin(\theta)|V\rangle$ to a time-bin qubit with an encoding converter. Then the time-bin qubit is further converted into a path qubit to be stored in two memory cells and finally read out and converted back into a polarization qubit $a_4|H\rangle+b_4|V\rangle=cos(\theta_4)|H\rangle+e^{i\phi_4}sin(\theta_4)|V\rangle$, which has a high overlap fidelity with the input state if $\theta_4=\theta$ and $\phi_4=\phi$. This can be achieved by calibrating the amplitude and phase in the read-out AODs. (d), The index of the $72$ qubit cells paired from 144 micro-ensembles. (e)-(f), State fidelity of the retrieved qubit in six typical qubit cells $1$, $4$, $20$, $22$, $34$ and $53$. Fidelities for different input polarizations with $15\mu$s storage are in (e), and average fidelities over $4$ polarizations at different storage time are in (f). The crosstalk error induced by accessing nearby micro-ensembles of the stored qubit. The green micro-ensemble pair represents the qubit for storage, and we iteratively access the $6$ neighboring regions marked grey for different rounds to characterize the fidelity loss. Note that the storage time is fixed to $55\,\mu$s with variable round number. A round of operations on the 6 neighbors leads to about $1\%$ infidelity. The error bars represent one standard deviation in (e), (f), and (g).}
\end{figure*}
\begin{figure*}
  \centering
  \includegraphics[width=17.4cm]{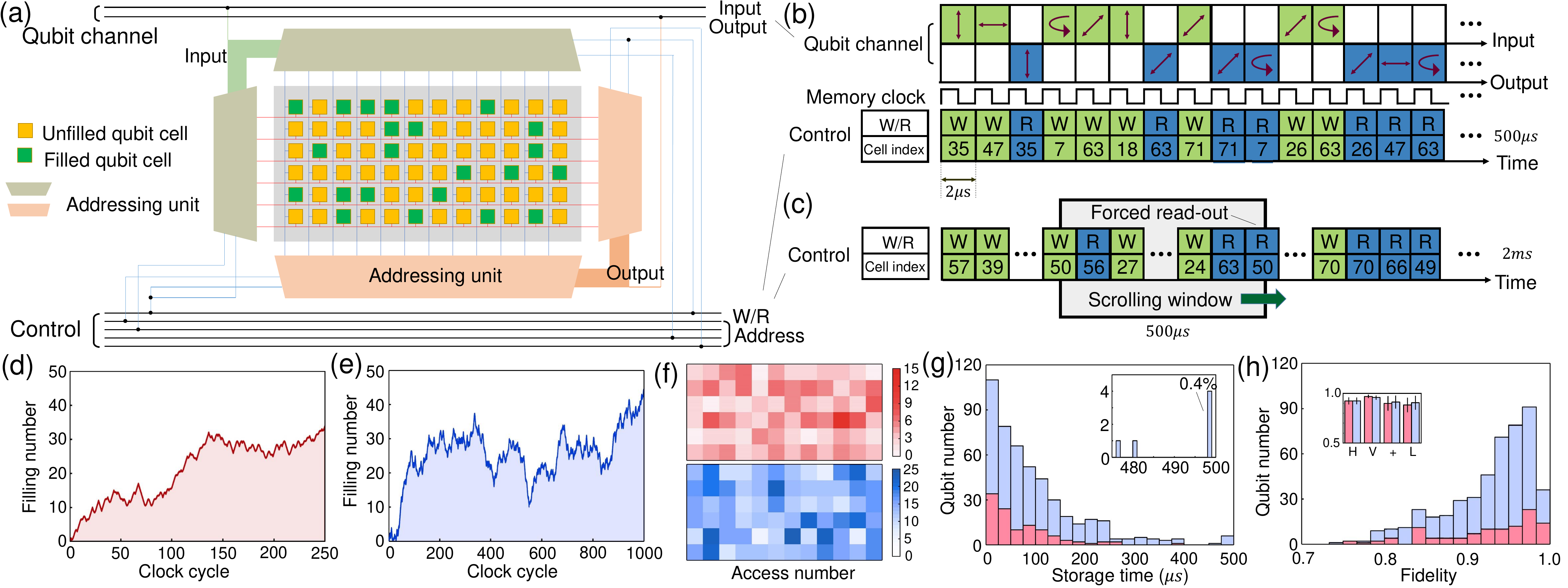}\\
  \caption{Random access quantum memory.
  (a) Snapshot of the RAQM with $72$ qubit cells. (b) A random sequence of $500\,\mu$s. All the domains in each $2\,\mu$s operation are randomly generated including write-or-read (W/R), qubit cell index, and input polarization. (c) A similar random sequence of $1000$ write or read operations with a length of $2\,$ms. We update the storage time of each stored qubit within a scrolling window of $500\,\mu$s, and execute forced read-out to the stored qubits about to expire in the next clock cycle. (d)-(h) The filling number of the memory, the total accesses in each qubit cell, the distribution of storage time of all output qubits, and the distribution of the state fidelities of all output qubits, for both $500\,\mu$s (red) and $2\,$ms (blue) cases. The inset in (g) shows the forced read-out happens at a chance of $0.4\%$. The inset in (h) demonstrates the corresponding fidelity for each input polarization. The error bars represent one standard deviation.
  }
\end{figure*}

\subsection{High-performance random access quantum memory}
The structure of the random access quantum memory is illustrated in Fig. 3(a). Two qubit channels are used for the input and output of photonic qubits. The classical instructions including write-or-read (W/R) and the index of qubit cell are sent to the addressing units via control channels, as shown in Fig. 3(a) and 3(b). Accomplishing a write or read operation for a qubit takes $2\,\mu$s to access a pair of micro-ensembles successively via upgraded optical design, which is about $10$ times faster in switching than the previous work~\cite{jiangnan}. Programming of this memory is to arbitrarily assign tasks in each $2\,\mu$s cycle of the memory clock, quite similar to piling up a Lego, with one condition that the stored qubit can't be read out multiple times due to the quantum non-cloning theorem \cite{non_cloning}.

We first characterize the performance of this RAQM with a randomly generated $500\,\mu$s (250 operations) sequence. Here we set the total length of the sequence within the coherence time to guarantee the faithful storage for each stored qubit. All the domains in each of the $250$ operations are randomly generated, including write or read (W/R), index of qubit cell, and the polarization for each input qubit, as illustrated in Fig.~3(b) and Fig.~12. Here the input qubit is carried by a weak coherent state ($\bar{n}=0.5$). We use several metrics to demonstrate the sequence and characterize the performance of the memory. The dynamic filling number representing the real-time number of stored qubits, the number of total accesses at each of the $72$ qubit cells, and the distribution of storage time and fidelity for each retrieved qubit are illustrated in Fig.~3(d)-3(h). In this sequence, the most frequently accessed cell is visited by $12$ times, compared to the average number of $3.74$ (Fig.~3(f)). The average storage time for all the $108$ retrieved qubits is about $68.6\,\mu$s. The fidelity distribution of the retrieved states is shown in Fig.~3(h), with  average fidelity of $93(3)\%$, $97(1)\%$, $90(7)\%$, and $88(7)\%$ for different input polarization $|H\rangle$, $|V\rangle$, $|+\rangle$, $|L\rangle$, respectively.

\begin{figure*}
  \centering
  \includegraphics[width=15cm]{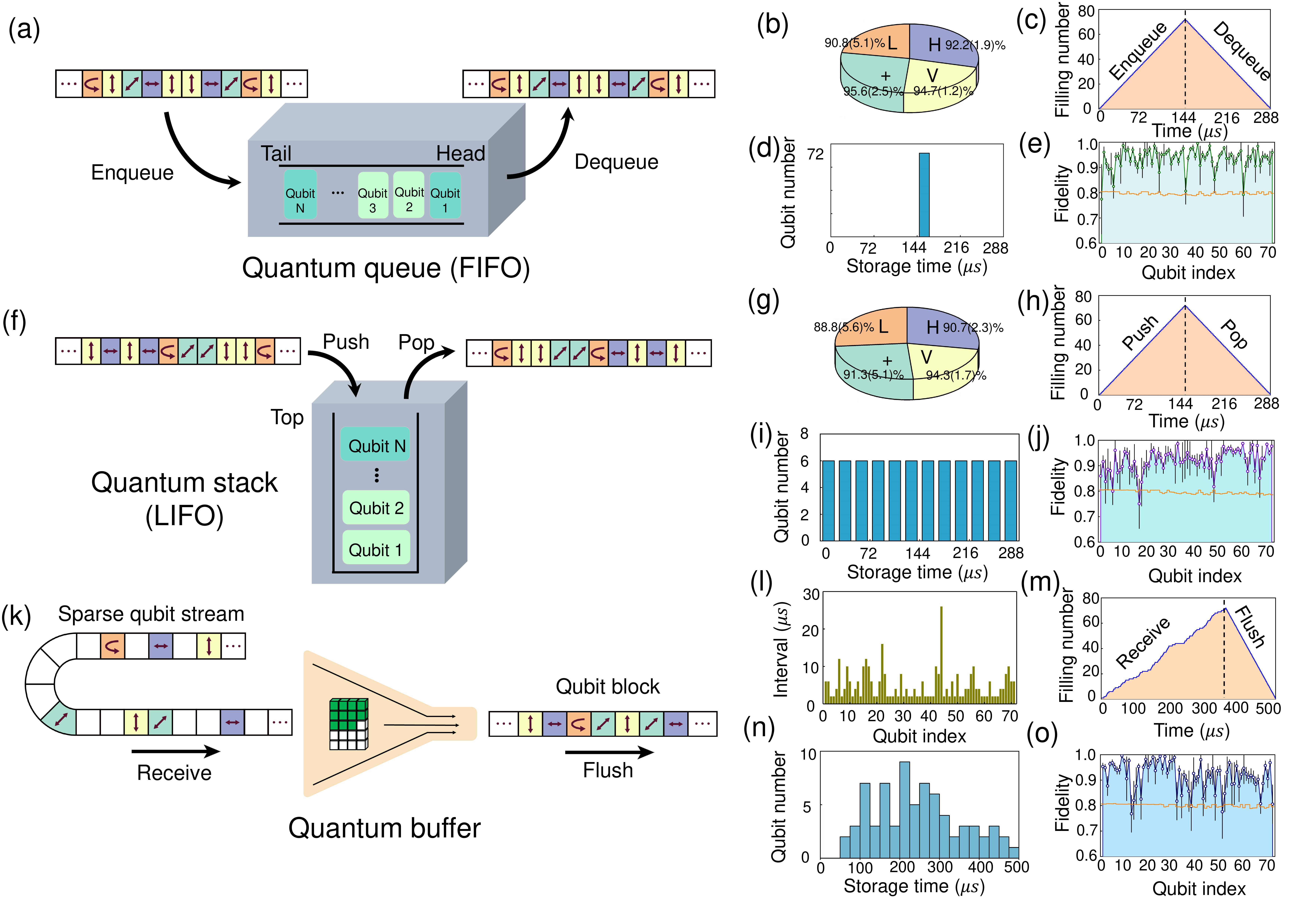}\\
  \caption{Quantum queue, stack, and buffer.
   (a) Quantum queue (FIFO). We demonstrate a special sequence of $72$ contiguous enqueues followed by $72$ contiguous dequeues under the rule of a queue. (b)-(e)  The polarization distribution of the input states and average fidelity for each input polarization, the real-time filling number, the distribution of storage time for individual qubits, and the fidelities for individual qubits and corresponding threshold for quantum storage (the orange lines) \cite{jiangnan}, respectively. (f)-(j) Schematic and results for a quantum stack. Again we use a special sequence in which we first push $72$ qubits and then pop them out in a reversed order. (k)-(o) Schematic and results for a quantum buffer. The distribution of time intervals between successive input qubits are shown in (l). The error bars represent one standard deviation in (b), (e), (g), (j), and (o). }
\end{figure*}

Limiting the running time within qubit coherence is a strong condition but not necessary. Instead, we only need to fulfill a weaker one that each in-memory qubit stays within the coherence time. To prevent the storage time from exceeding the coherence time, we maintain a scrolling window of $500\,\mu$s which keeps track of the storage time of each occupied qubit cell inside the window, and force the quantum memory to execute read-out if a stored qubit is about to reach the maximum storage time of $500\,\mu$s in the next cycle. In this way we further elongate the sequence to $2\,$ms, corresponding to a total number of $1000$ continuous operations in one run, which is almost two orders of magnitude larger than the previous record~\cite{rempe_ram, wolters}. We characterize the performance of this memory with a randomly generated sequence.  Totally $478$ output qubits are read out, with average fidelity of $92(3)\%$, $96(2)\%$, $91(6)\%$, and $90(7)\%$ for each input polarization $|H\rangle$, $|V\rangle$, $|+\rangle$, and $|L\rangle$. Each qubit cell is accessed by $13.9$ times on average  (Fig.~3(f)). We find the forced read-out happens at a quite rare chance below $1\%$ (Fig.~3(g)), which has negligible influence on the programming of sequence or the performance of memory. More details regarding the sequences and results can be found in the Appendices.

The fidelities of $|+\rangle$ and $|L\rangle$ are slightly lower than the fidelities of $|H\rangle$ and $|V\rangle$ as they suffer from the phase uncertainty in the encoding converter due to imperfect interferometer locking but $|H\rangle$ and $|V\rangle$ don't. The higher fidelity in $|V\rangle$ versus $|H\rangle$ mainly origins from the slight mismatch of the memory array center with the center of the atomic cloud, which induce higher efficiency in the $|V\rangle$ basis than the $|H\rangle$ basis thus better signal-to-noise ratio.

\begin{figure*}
  \centering
  \includegraphics[width=15cm]{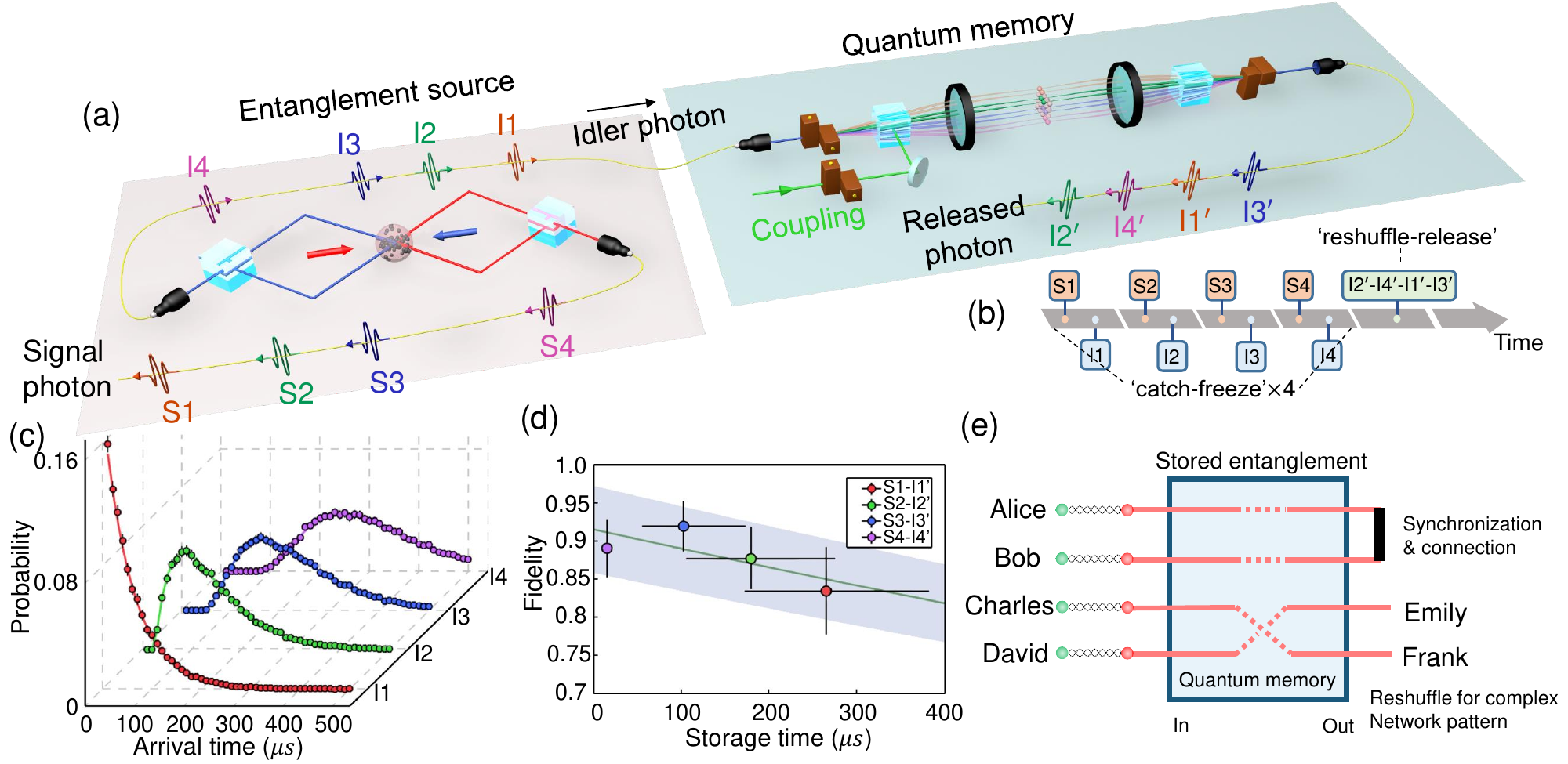}\\
  \caption{Storage and reshuffle of multiple entangled photon pairs.  (a) Four Einstein-Podolsky-Rosen (EPR) pairs are sequentially generated through the DLCZ protocol and are heralded by the detection of signal photons. The $4$ idler photons with random arrival times are stored into the quantum memory and are read out together in an arbitrarily chosen order ($2$-$4$-$1$-$3$). (b) The sequence of the experiment. (c) The distribution of the random arrival times of the four idler photons $I_1$, $I_2$, $I_3$, and $I_4$. Note that the four EPR pairs are generated sequentially so that the distributions are not independent. The error bars represent one standard deviation. (d) The entanglement fidelity of the four EPR pairs $S_1$-$I_1'$, $S_2$-$I_2'$, $S_3$-$I_3'$, and $S_4$-$I_4'$ after storage and reshuffle. The fit follows an exponential decay with the gray shaded area representing the $68$\% confidence interval. The error bars also represent $68$\% confidence interval. (e) Two potential applications in a quantum network, i.e. entanglement connection of two quantum repeater segments and entanglement routing between different network nodes.}
\end{figure*}

\subsection{Quantum queue, stack, and buffer}
A queue, also known as a first-in-first-out (FIFO) data structure, follows a special operation rule that the input (enqueue) and output (dequeue) of data can only happen at the tail or head of the queue, as illustrated in Fig.~4(a). With modifications in the control logic, we implement a quantum queue based on the RAQM, to receive and release photonic qubits in a first-in-first-out style. Unlike the sequences in the RAQM where the inputs and outputs are intertwined, here we choose to demonstrate a special case in which $72$ continuous enqueues are first performed to fully fill the $72$ qubit cells, followed by $72$ consecutive dequeues, to demonstrate the feature of FIFO maximally, as illustrated in Fig.~4(c) and Fig.~14. In each clock cycle of $2\,\mu$s, the input qubit is chosen from $|H\rangle$, $|V\rangle$, $|+\rangle$, $|L\rangle$ with uniform probability. In this special sequence, all the $72$ input qubits are read out in the end and each of them has the same storage time of $144\,\mu$s, due to the rule of FIFO (Fig.~4(d)). The fidelities of individual input qubits are listed in Fig.~4(e), and the average fidelity for each input polarization is illustrated in Fig.~4(b).

As a counterpart to a queue, a stack is another abstract data structure which has only one data entrance (top), for both input (push) and output (pop), as shown in Fig. 4(f). In a stack, the most remarkable property is that the input data received later will be output earlier, thus a stack is also known as a last-in-first-out (LIFO) data structure. Here we still employ a special sequence in which $72$ qubits are first pushed in a row, and later popped out continuously. A general sequence with pushes and pops interweaved is described in detail in Fig.~17. In this case we present here, several statistics and measured results are illustrated in Fig.~4(g)-4(j). The storage time of the $72$ output qubits follows a uniform distribution ranging from $2\,\mu$s to $286\,\mu$s, in sharp contrast to the constant $144\,\mu$s for the queue, as a result of the FIFO and LIFO features of these two data structures. The fidelities of the $72$ output photonic qubits are shown in Fig.~4(j). We note that there is slight fidelity decay for the earlier pushed qubits due to longer storage time in the stack.

Buffer, a memory which usually receives data at one rate and releases them at another rate, is often used to bridge different components in the network, such as online video streaming and distributed computing in classical information technology. In analogy to the classical buffer, here we demonstrate a quantum buffer which can receive qubits from a sparse stream, synchronize them, and then flush all qubits out as a block in an arbitrary order, as illustrated in Fig.~4(k). The sequence for characterizing the buffer is generated with random arrival times and input polarizations. We can clearly see the rate difference in the $356\,\mu$s receiving stage and the following $144\,\mu$s flushing stage in Fig.~4(m).  The average fidelity for each input polarization is demonstrated in Fig.~15(b).

Here we find a couple of qubits in the $72$-qubit sequences may fall below the threshold of quantum storage~\cite{jiangnan}, as illustrated in Fig.~4(e),~4(j), 4(o), 16(e), and 17(e). These occasional low-fidelity points are mainly induced by imperfect calibration in the experiment, which doesn't influence the quantum nature of the corresponding qubit cells. A detailed discussion about this issue is in Appendix E.

\subsection{Storage and reshuffle of multiple quantum entanglements}
Finally, we demonstrate the catch and storage  of 4 heralded entanglements from a probabilistic entanglement source, and later the release of them in an arbitrary order regardless of the incoming sequence, through the programmable multi-purpose quantum memory implemented here. It is noteworthy that the combination of an entanglement source and a write-read quantum memory is sufficient for heralded entanglement generation~\cite{1250,1650,rmp_gisin}, which serves as one of the core functions in a quantum repeater.

The entanglement source is based on the Duan-Lukin-Cirac-Zoller (DLCZ) protocol \cite{DLCZ}, in which repeated write-read laser pulses probabilistically excite an entangled state between a signal photon and an idler photon $|\psi\rangle=\frac{|H_sV_i\rangle+|V_sH_i\rangle}{\sqrt{2}}$, as illustrated in Fig.~5(a) and the Appendices. Here we can receive an entangled signal-idler pair from the source in about $70\,\mu$s, and $4$ signal-idler entangled pair in $500\,\mu$s with high probability. The fidelity of the signal-idler entangled pair from the source is measured to be $94(1)\%$ via quantum state tomography.

Unlike the weak coherent state used in the RAQM and quantum queue, stack, and buffer, here the quantum information is carried by single photons. As illustrated in Fig. 5(a) and 5(b), the protocol begins with repeated excitation of signal-idler entangled pair. Upon a success is heralded by the detection of a signal photon, the control unit dynamically prepares the quantum memory for the incoming idler photon and stores it into a qubit cell. We call this `catch' and `freeze' in a vivid way. We repeat `catch-freeze' $4$ times to store the idler photons from $4$ randomly generated entangled pairs $S_1$-$I_1$, $\dots$ $S_4$-$I_4$ into the memory. The measured distribution of arrival times for the $4$ idler photons is illustrated in Fig.~5(c). After all the $4$ idler photons are stored in the quantum memory, the control unit steers the memory to read out all the stored qubits in a row ($4$ consecutive clock cycles) with any desired order, and we call this `reshuffle' and `release'. Here we demonstrate a reshuffled order of $2$-$4$-$1$-$3$, which means the second arrived idler photon from the entanglement source will be first read out from the memory in the final `release' stage, and so on. Therefore, the final entangled pairs are $S_2$-$I_2'$, $S_4$-$I_4'$, $S_1$-$I_1'$, and $S_3$-$I_3'$, sorted by idler photon release time, as shown in Fig. 5(a) and 5(b). We measure the fidelities for these reshuffled entangled photon pairs and the distribution of the storage time for each pair, as shown in Fig. 5(d). The fidelities of the $4$ entangled pairs after `catch-freeze-reshuffle-release' are between $83(6)\%$ and $92(3)\%$, with slight decay due to the increasing storage time.

Here we also present two potential applications via entanglement storage and reshuffle in a quantum network, as illustrated in Fig.~5(e) (Note that these two applications haven't been implemented in this work.). In the first potential application, Alice and Bob each create an entangled photon pair and send one photon to the quantum memory. The memory synchronizes~\cite{synchronization} these two photons with different arrival times and performs Bell state measurement between them to further entangle Alice and Bob. This resembles the entanglement connection \cite{lattice, lanyon_connection} between two repeater segments, which is also a core function in a quantum repeater. A possible method to output two photons at the same time is proposed in Appendix G.

In the second potential application, Charles and David want to share their entanglements with Frank and Emily respectively, however they only have quantum channels to an intermediate quantum memory. Thus Charles and David each send one photon from their entangled photon pair to the memory. The memory receives two photons, stores them, and sends the photons to their destinations by reading them out at appropriate clock cycles when Frank and Emily are ready to receive respectively, as illustrated in Fig.~5(e). In this potential application, the quantum memory can serve as a router which establishes quantum links inside a group of quantum nodes with full connectivity and substantially saved resources such as quantum channels.\\

\section{Discussion}
\noindent In summary, we demonstrate a high-performance multipurpose quantum memory which can be adapted for various core applications in a quantum network. The quantum memory presented in this work provides a promising platform for the implementation of quantum repeaters,  quantum network, and networked quantum computation, considering the optical qubit processing ability demonstrated in this work. In the future, it is possible to further improve the coherence time to over $0.1$s \cite{bao0.1s, 1min} and efficiency above $0.5$ \cite{zhushiliang, hsiao} with much higher optical depth by loading the atoms into an optical-lattice array, which will greatly improve the performance of photonic quantum logic \cite{klm}. We can also exploit wavelength conversion techniques \cite{changwei, ikuta, weifurner33km, lanyon_connection} to achieve a quantum repeater network for long-distance communication. Finally, if in-memory quantum gates are further equipped via Rydberg interactions \cite{rydberg} or spin superexchange \cite{yangbing}, a nearly universal platform can be realized which will support the implementation of fully-connected quantum logic, error-corrected quantum network, or even a global quantum internet \cite{kimble,weiner,repeater in space,global network}.

\section{Acknowledgements}
The authors thank Pengcheng Lai and Zuqing Wang for helpful discussions. This work is supported by Innovation Program for Quantum Science and Technology (No.2021ZD0301102), the Tsinghua University Initiative Scientific Research Program and the Ministry of Education of China through its fund to the IIIS. Y.F.P. acknowledges support from the start-up fund and the Dushi Program from Tsinghua University. Y.K.W. acknowledges support from the start-up fund from Tsinghua University and National Key Research and Development Program of China (2020YFA0309500).

\appendix

\section{Experimental methods}
Here we describe the experimental realization of a long-lived spatially-multiplexed quantum memory array with $144$ individually addressable microscopic ensembles in detail. In the beginning, we load a macroscopic $^{87}$Rb atomic cloud in a vacuum chamber by Magneto-optical trap (MOT). The macroscopic atomic ensemble consists of about $10^9$ atoms and the temperature is estimated to be $300\,\mu$K. We further apply molasses cooling (sub-Doppler cooling) for $1\,$ms to reduce the temperature to $20\,\mu$K. We use a two-dimensional addressing unit (AOD, DTSXY-400) to access different regions of the macroscopic atomic cloud. The AOD scanning frequency ranges from $85\,$MHz to $118\,$MHz with a step of $3\,$MHz in both X and Y directions, as illustrated in Fig. 6. The step of $3\,$MHz is chosen based on a tradeoff between faster access time and larger memory capacity, which will be further explained in Appendix D. By tuning the RF frequencies on the AOD, one can access each of the $12\times 12=144$ microscopic ensembles with a constant time of $1\,\mu$s. Every two adjacent microscopic ensembles are paired to constitute a qubit cell, as shown in Fig. 2(d).

\begin{figure}
  \centering
  \includegraphics[width=8cm]{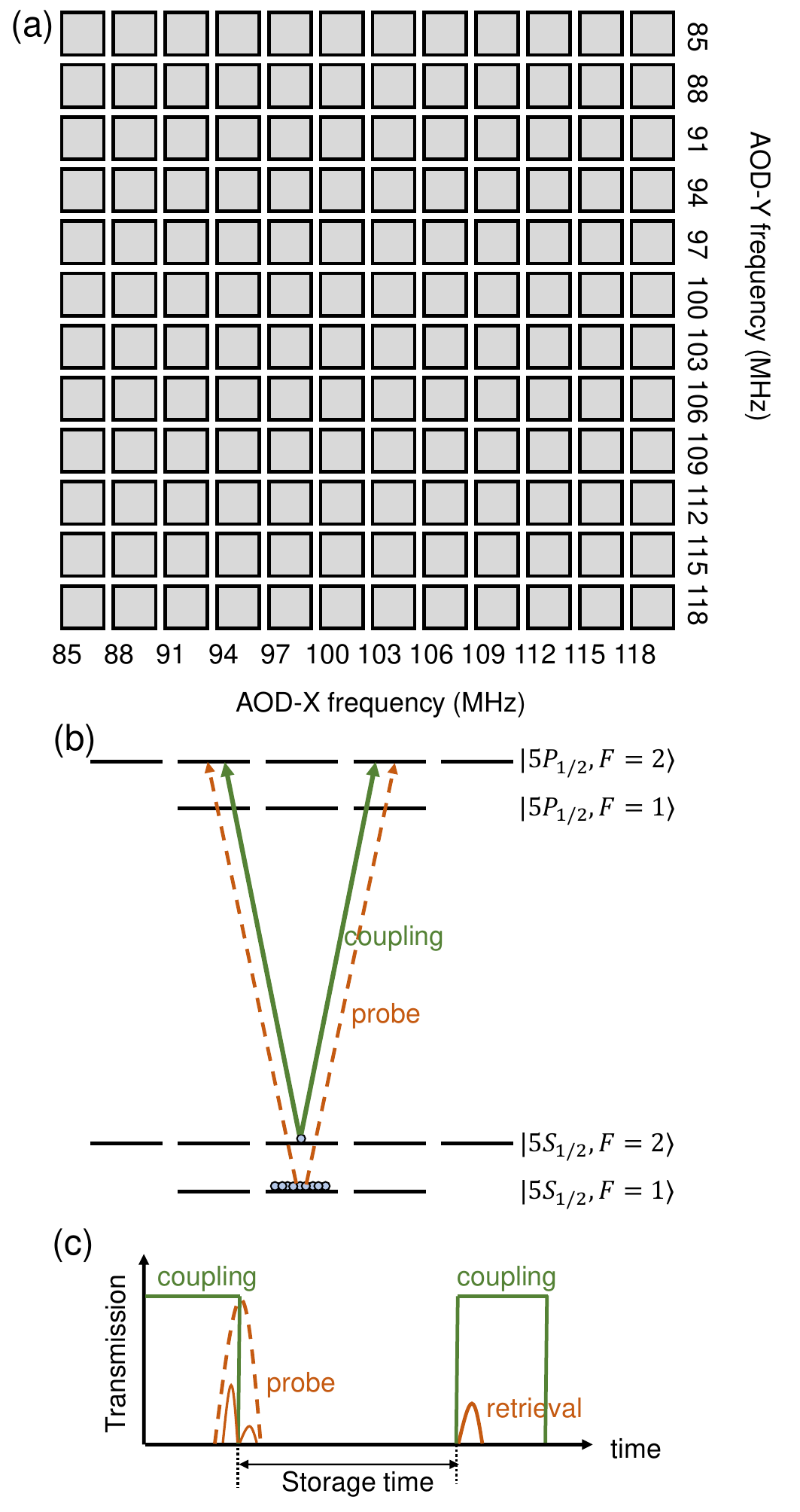}\\
  \caption{Quantum memory array and EIT storage. (a) The two-dimensional quantum memory array with 144 microscopic ensembles. The AOD frequencies for X and Y dimensions corresponding to each microscopic ensemble are marked in the figure. (b) The energy level configuration for the EIT process. (c) A schematic for the photon write-in process and read-out process through EIT.
  }
\end{figure}

\begin{figure}
  \centering
  \includegraphics[width=8.7cm]{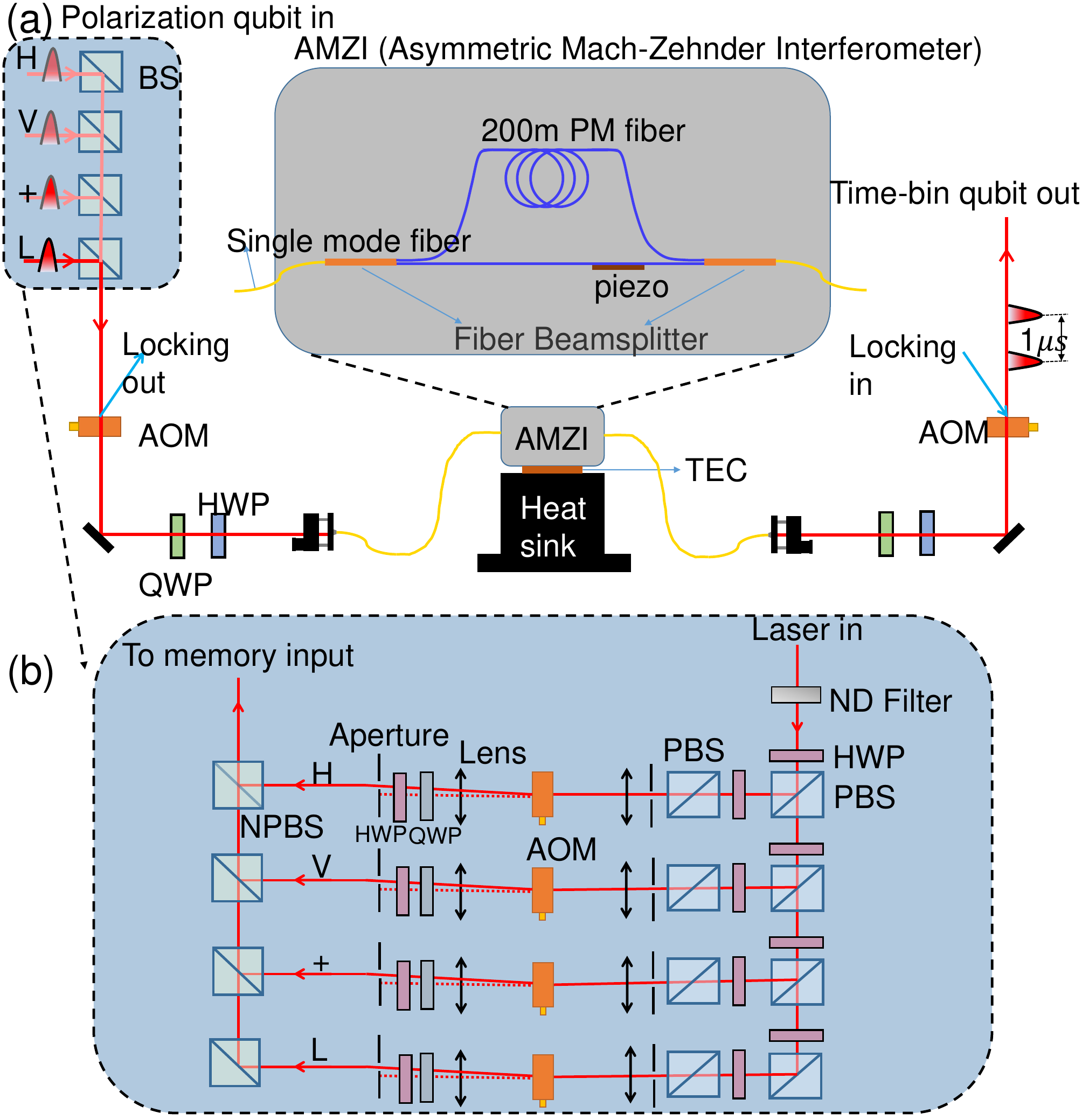}\\
  \caption{The asymmetric Mach-Zehnder interferometer and fast switch network for input state preparation. (a) The AMZI is temperature-stabilized to avoid phase drift. It is sealed in an aluminum box. The half-wave plate (HWP) and quarter-wave plate (QWP) are inserted to compensate for the polarization change in fiber transmission. The fiber beamsplitter is PBC780SM-APC (Thorlabs). (b) The fast switch network for different input states. Here we first split the input laser into four different paths as different polarization input states, and then control the on/off of each path by controlling four different acoustic optical modulators (AOMs) as fast switches. Note that in each clock cycle, no more than one AOM should be turned on. Then we combine the four input coherent states by a NPBS (non-polarizing beamsplitter) and send it into the input channel of the memory. Each polarization can be carefully adjusted by a pair of HWP and QWP before the NPBS.
  }
  \end{figure}

\section{Long coherence time}
The decoherence in the atomic-ensemble-based quantum memory is mainly induced by atomic random motion and magnetic field noise. To suppress the magnetic field influence, we employ the clock transition between $|g\rangle=|5S_{1/2},F=1,m=0\rangle$ and $|s\rangle=|5S_{1/2},F=2,m=0\rangle$. The initial state $|g\rangle$ is prepared through a $20\,\mu$s optical pumping process in which two $\sigma$ lasers and one $\pi$ laser are used. The two $\sigma$ lasers are resonant to $\ket{5S_{1/2},F=2}\rightarrow\ket{5P_{1/2},F=2}$ transition and $\ket{5S_{1/2},F=2}\rightarrow\ket{5P_{3/2},F=2}$ transition, respectively. The $\pi$ laser is resonant to $\ket{5S_{1/2},F=1}\rightarrow\ket{5P_{1/2},F=1}$ transition. We employ a collinear beam configuration to suppress the effect of atomic random motion and use three cascaded etalons to filter out noise. The transmission efficiency for signal photon is $90\%$ in each etalon while the coupling light is suppressed by $500$ times. The probe light is H polarized and the coupling light is V polarized. The probe (coupling) light focuses on the center of the atomic micro-ensemble with a beam waist of $125\,$($160$)$\,\mu$m. The energy level for the EIT process is shown in Fig.~6(b). With all these efforts, we extend the storage time of each microscopic ensemble to $500\,\mu$s level.

\begin{figure}
  \centering
  \includegraphics[width=8.7cm]{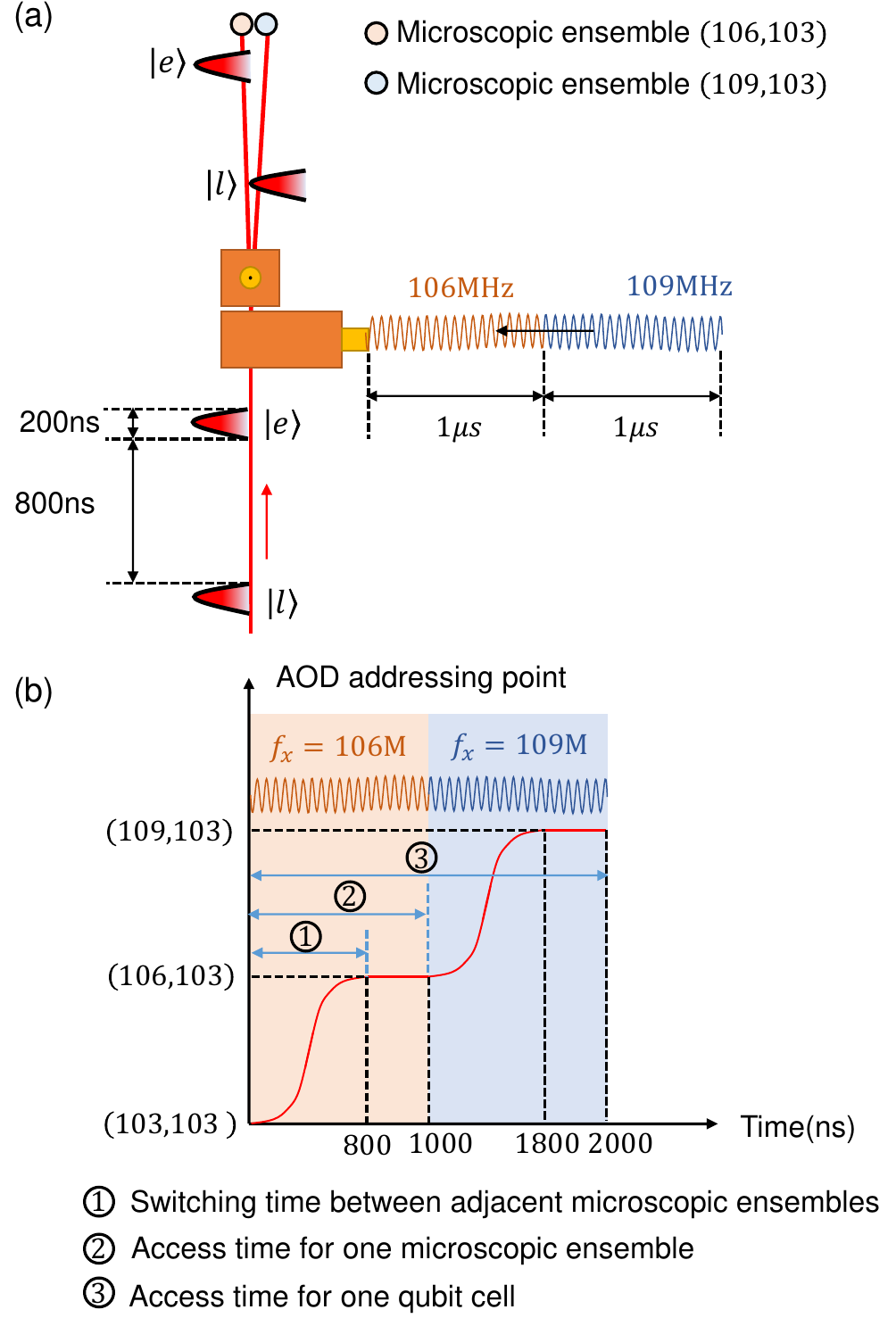}
  \caption{Storage of a time-bin qubit into two adjacent microscopic ensembles. (a) The schematic for AOD addressing. The RF signals are generated by the arbitrary waveform generators and go through a switch and an amplifier to drive the AOD. (b) The move of the AOD addressing point as RF signals of different frequencies are fed into the AOD. Every time we visit a microscopic ensemble, the AOD addressing point keeps stationary for $200\,$ns. During the $200\,$ns, the EIT process happens and the photons are stored in or retrieved from the memory.
  }
\end{figure}

\begin{figure}
  \centering
  \includegraphics[width=8.7cm]{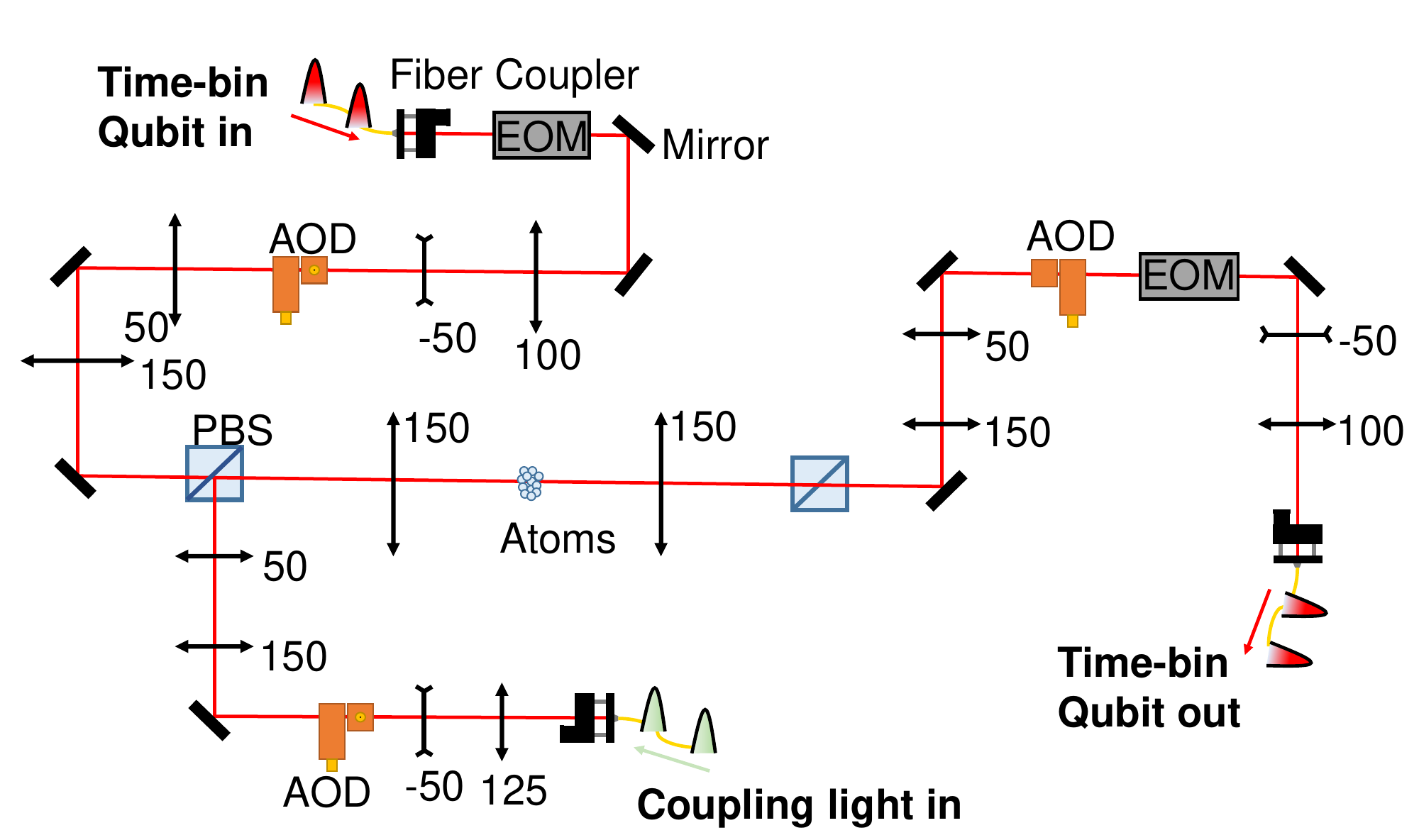}
  \caption{Optical configurations in detail. We use multiple sets of telescopes that adjust the beam waist, in front of and behind each AOD. The units of all the focal lengths are in mm. The telescope system in front of the AOD ($100\,$mm and $-50\,$mm) is to narrow the beam waist for faster switching time on the AOD. The telescope system behind the AOD ($50\,$mm and $150\,$mm) is to enlarge the beam waist to obtain a small focus radius on the atoms. The $150\,$mm lenses on the left and right sides of the atomic ensemble constitute a 4f system. The design of the coupling light path is similar to that of the signal light, the only difference is the magnification factor of the telescope. The encoding converters are not shown for simplicity.}
\end{figure}

\section{Encoding conversion }
Polarization is one of the most widely used encodings for a photonic qubit. However, it is not straightforward for our multiplexed memory array to store polarization qubits because the addressing unit (AOD) is polarization-selective. One solution to this problem is to convert the polarization encoding to time-bin encoding. As shown in Fig. 7(a), we design an asymmetric Mach-Zehnder interferometer (AMZI) to carry out this task \cite{2}. The AMZI consists of two fiber beamsplitters and two polarization-maintaining (PM) single-mode fibers with a path difference of $200\,$m. The path difference of $200\,$m corresponds to the time interval of $1\,\mu$s for the time-bin qubit. 1$\,\mu$s is also the access time for each microscopic ensemble. The interferometer is integrated in an aluminum box and the temperature is stabilised with a fluctuation less than $0.004^{\circ}$C. The path difference of the AMZI is actively locked by stretching the fiber stuck to a piezo. The locking laser is the $+1$st order diffraction of the acousto-optic modulator (AOM) and is coupled into the AMZI. Finally, it comes out from the $+1$st order of the AOM on the other side. During the memory operation, the AMZI locking is paused, the locking laser is temporally turned off, and the signal light passes directly through the zero order of the two AOMs. The servo of the AMZI works during the MOT loading and molasses cooling stage.

We achieve the fast preparation of weak coherent pulses for different input states via a switch network. The switch network consists of 4 acoustic optical modulators (AOMs), each for turning on/off one of the four polarization inputs, as well as several beamsplitters and waveplates for beam combining and polarization adjustments (Fig.~7(b)). The turn-on delay of the AOM is less than $200\,$ns and the rise/fall time is $30\,$ns, which is fast enough compared to the AOD switching time ($\sim1\,\mu$s).

 We use an arbitrary input state $a|H\rangle+b|V\rangle$ as an example to illustrate the addressing and storage process. As shown in Fig. 2(c), if an arbitrary input polarization $a|H\rangle+b|V\rangle =cos(\theta)|H\rangle+e^{i\phi}sin(\theta)|V\rangle$ enters the memory, it is first converted to a time-bin qubit $a_1|E\rangle+b_1|L\rangle= cos(\theta_1)|E\rangle+e^{i\phi_1}sin(\theta_1)|L\rangle$ at the input encoding converter. Here($\theta,\phi$) and ($\theta_1,\phi_1$) can be different as the efficiency imbalance and a constant phase offset induced by the encoding converter. Then the AOD sequentially deflects $|E\rangle$ to $|U\rangle$ and $|L\rangle$ to $|D\rangle$, and stores the time-bin qubit into the atomic ensemble as a spatial qubit $a_2|U\rangle+b_2|D\rangle$ after two successive EIT storages. After some programmable storage time, the stored qubit is read out and converted back to time-bin encoding $a_3|E\rangle+b_3|L\rangle$ again by AODs and EIT, as a reversed step in EIT storage. Finally, the qubit is converted back to a polarization qubit   $a_4|H\rangle+b_4|V\rangle=cos(\theta_4)|H\rangle+e^{i\phi_4}sin(\theta_4)|V\rangle$ after the output encoding converter. One can see that, if ($\theta_4,\phi_4$) is very close to ($\theta,\phi$), we can guarantee that the output state has a high fidelity with the input state. In this work this is achieved by carefully tuning efficiency imbalance and phase in the two paths ($|H\rangle\rightarrow|E\rangle\rightarrow|U\rangle\rightarrow|E\rangle\rightarrow|H\rangle$ and $|V\rangle\rightarrow|L\rangle\rightarrow|D\rangle\rightarrow|L\rangle\rightarrow|V\rangle$) via tuning the RF signals sent to AODs.

\section{Fast switching between different microscopic ensembles and memory clock}
In this section, we discuss the switching time between different microscopic ensembles. The switching time results from the time for the sound wave to pass through the laser area of the AOD. Roughly we have the following relationship
\begin{equation}
  \text{switching time}\approx\frac{w}{v_s},
\end{equation}
where $w$ is the laser beam waist on the AOD, and $v_s$ is the velocity of the sound wave in the AOD crystal. Since $v_s$ is determined by material ($650\,$m/s), we can only reduce the light beam diameter on the AOD. However, the light beam diameter can not be too small because a smaller diameter on the AOD results in fewer resolvable points on the macroscopic atomic ensemble as well as lower AOD diffraction efficiency. Given this trade-off, we set the switching time between the adjacent microscopic ensembles to $800\,$ns and use the following $200\,$ns for qubit storage or retrieval. As shown in Fig. 8(a) and 8(b), the access time is $1\,\mu$s for one microscopic ensemble and thus $2\,\mu$s for one qubit cell. The scanning frequency step from one microscopic ensemble to its nearest neighbor is $3\,$MHz in both X and Y directions. The maximally allowed addressing capacity is $12\times 12=144$ microscopic ensembles. The optical configurations are shown in Fig. 9.

\section{Sources of crosstalk and infidelity}

The crosstalk in this experiment origins from three sources. The first and dominating source is the leakage light of coupling pulse during the operation on the neighbor cells, which could ruin the stored quantum state in the current cell. This source of crosstalk is characterized by the linear decrease in the fidelity when the number of operations (to deliberately induce crosstalk by operating the surrounding cells) increases as illustrated in Fig. 2(g). Note that the storage time is fixed when we increased the number of rounds in Fig. 2(g), thus the fidelity decay due to decoherence can be ruled out. We find that one round of operations on the 6 neighbors leads to about $1\%$ infidelity due to this crosstalk source.

The center-to-center distance between two micro-ensembles is $190\,\mu$m, the $1/e^2$ radius of the coupling beam is $80\,\mu$m, and the $1/e^2$ radius of the input/output photonic modes (equivalent to the size of each micro-ensemble) is $63\,\mu$m.

The second source of crosstalk is the atomic diffusion. This influence will become significant if the storage time reaches one millisecond, but has negligible influence in the current experiment considering the typical storage time in our experiment is about $200\,\mu$s (see Fig.~3(g)). As the temperature of the atomic cloud is $20\,\mu$K in our experiment, we estimate the $1/e^2$ radius of the micro-ensemble increases slightly from $63\,\mu$m to $65\,\mu$m after $200\,\mu$s of atomic thermal motion, which is not enough to induce observable crosstalk as the center-to-center separation between two modes is $190\,\mu$m.

The third source is the drop of the micro-ensemble due to gravity. As we did not load the micro-ensembles into optical traps like in the tweezer array experiments, the micro-ensembles experience free fall during the storage. This crosstalk source is even less significant than the atomic thermal motion under current situation, as the displacement due to gravity is only $5\,\mu$m after $1\,$ms of free fall. This source of crosstalk will become significant when the storage time reaches several milliseconds.

Here we also discuss the sources of the low-fidelity storage in a long sequence, as shown in Fig.~4(e),~4(j), 4(o), 16(e), and 17(e). These low fidelity cases are mainly induced by the imperfect calibration for a long sequence. In all the measurement results in our earlier work~\cite{jiangnan} and Fig.~2 in this work, the storage fidelity of each qubit is measured separately (i.e. only one qubit is measured in a single measurement, and careful calibration by hand is applied before each measurement is taken). Thus careful calibration including the relative phase compensation of the two paths and the efficiency balance (see Fig. 2(c)) can be applied to the AOD addressing unit to optimize the storage fidelity. However in the measurement results shown in Fig. 3, 4, 5, 16, 17 in this manuscript, the fidelity for all the qubits in the whole sequence are measured in one continuous run, which is much more advanced than our previous work. However in this case, careful calibration for each qubit cell becomes very difficult, as it is now very time-consuming to adjust each cell by hand. Thus we only did a fast and rough calibration of each cell before measurement, which can induce occasional low-fidelity cases. If each qubit cell is carefully calibrated, there is no doubt all the qubits can surpass the quantum bound. For example, the first qubit in Fig. 4(e) which has a fidelity of $77.5\pm13.8\%$ is stored in qubit cell no.1, and this point fails to surpass the bound. However when we measure this cell separately in Fig. 2(e), the fidelities for all polarizations are over $90\%$ and all above the bound. In the future, we could switch the calibration from manual to automatic for better calibration. Other infidelity sources include crosstalk, storage time, and inhomogeneity of memory efficiency, and all of these experimental imperfections need to be improved to reduce the probability of low fidelity cases.

 \begin{figure}
  \centering
  \includegraphics[width=8.7cm]{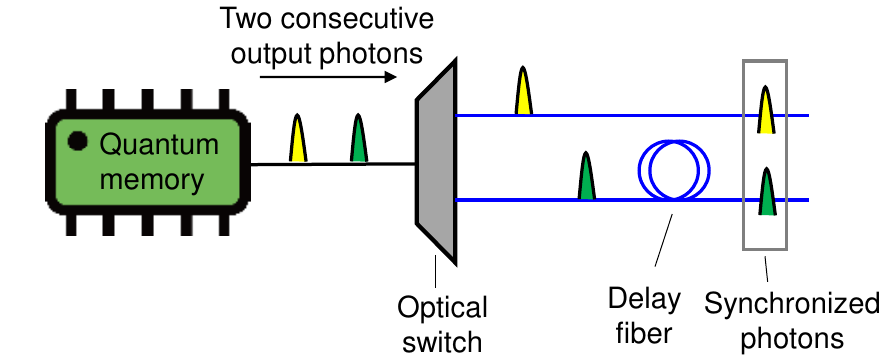}
  \caption{Potential method for simultaneously output two photons. As our multi-purpose quantum memory has only one output channel, we can't output two stored photons at a time. Here we propose to add a combination of an optical switch and a delayed fiber, in order to solve this problem. The memory first outputs two photons consecutively, with the green photon $2\,\mu$s ahead of the yellow photon. Then the optical switch directs the green photon to the lower path, and $2\,\mu$s later directs the yellow photon to the upper path. Due to the delay fiber, the lower path is $\sim400\,$m (corresponds to $\sim2\,\mu$s delay) longer than the upper path, thus the two photons exit the two paths simultaneously.}
\end{figure}

\renewcommand{\arraystretch}{1.5}
\renewcommand\tabularxcolumn[1]{m{#1}}
\begin{table}[h!]
  \begin{center}
  \small
    \begin{tabularx}{0.5\textwidth} {
    >{\centering\arraybackslash}X
    >{\centering\arraybackslash}X
    >{\centering\arraybackslash}X
    >{\centering\arraybackslash}X
    >{\centering\arraybackslash}X
    >{\centering\arraybackslash}X }
    \hline
    \hline
    Components &  Efficiency/transmission   \\[1pt]
    \hline
    Input fiber coupling  &  $85\%$ \\
    Input encoding converter (including AMZI, EOM and locking AOM) & $51\%$ \\
    Input AOD pair & $85\%$ \\
    Storage and retrieval in atoms (on average) &  $5.5\%$\\
    Output AOD pair &  $85\%$  \\
    Output encoding converter (including AMZI, EOM and locking AOM)  &  $52\%$\\
    Output fiber coupling & $85\%$\\
    Three Filter etalons & $73\%$\\
    Other optical elements transmission (mirrors, lenses, vacuum chamber window, etc.) & $90\%$\\
    \hline
    End-to-end efficiency & $0.50\%$\\
    \hline
    \hline
    \end{tabularx}
  \end{center}
  \caption{List of efficiencies for all  components in the write and read of an optical qubit.}
\end{table}

\section{Limiting factors for storage time and improvement strategies}

The dominating limiting factors for storage time in the current experiment is the atomic diffusion due to thermal motion and free fall due to gravity. Due to these two factors, the atoms in the spin-wave will leave the spatial extent of the signal mode and compromise the readout. The typical time-scale for these two factors is about a couple of milliseconds.

In our mind, the first measure to improve the storage time in the future should be holding the microensembles with optical lattice array. With optical lattice holding the atomic ensemble and applying magic-valued magnetic field which can cancel the differential AC stark shift in the optical trap, the atomic thermal diffusion and free fall can be significantly suppressed and the storage time would be improved from $\sim1\,$ms (in this experiment) to $\sim100\,$ms \cite{kuzmich0.1s}. Our plan would be to first load the atoms into a 2D optical trap array \cite{ensemble array} to solve the problems of atom diffusion and gravity, and see whether we can extend the storage time of each cell to about $100\,$ms.

In the second step we would try dynamical decoupling or spin echo to further extend the storage time to beyond one second. However, it is difficult to apply dynamical decoupling to an microensemble array via rf-pulses, due to the potential inhomogeneity of the rf amplitude across the large array (the spatial extent of the whole array is $2-3\,$mm in our case) and lack of the ability to individually address each cell (without influencing other cells). Although there was an experimental realization to use dynamical decoupling based on rf-pulses to extend the storage time of an atomic ensemble quantum memory to $16$ seconds \cite{1min}, how to adapt this method to a memory array with larger size and enable individual control of each microensemble across the array is not an easy task. A possible solution to this problem could be replacing rf with laser. The dynamic decoupling pulses can also be implemented via Raman transition \cite{Raman} with a small beam size which can individually control each cell to avoid crosstalks. On the other side, the laser-based gates usually have lower fidelities than the rf-based gates, which can be a problem if the number of flips goes large in the dynamic decoupling sequences. Replacing the Raman with STIRAP (Stimulated Raman adiabatic passage) may be a good idea to mitigate this problem.

\begin{algorithm*}
\begin{small}
\caption{~RAQM sequence ($500\,\mu$s)}\label{alg:cap}
\begin{algorithmic}
\State{\textbf{Data}: operation, qubit, address}
\State filling $\gets 0$, $i \gets 1$
\While{$i \le  250$}
\If{filling $= 0$}
    \State operation[$i$] $\gets$ write
    \State address[$i$] $\gets$ a qubit cell randomly selected from $1$ to $72$
    \State qubit[$i$] $\gets$ a polarization randomly selected from $\{\ket{H},\ket{V},\ket{+},\ket{L}\}$
    \State filling $\gets$ filling$+1$
\ElsIf{filling $=72$}
    \State operation[$i$] $\gets$ read
    \State address[$i$] $\gets$ a qubit cell randomly selected from $1$ to $72$
    \State filling  $\gets$  filling$-1$
\Else
    \State operation[$i$] $\gets$ write or read randomly determined by a probability distribution specified in Fig. 11
\If{operation[$i$] $=$ write}
    \State address[$i$] $\gets$ a qubit cell randomly selected from the empty cells
    \State qubit[$i$] $\gets$ randomly selected from $\{\ket{H},\ket{V},\ket{+},\ket{L}\}$
    \State filling $\gets$ filling$+1$
\Else
    \State address[$i$] $\gets$ a qubit cell randomly selected from the occupied cells
    \State filling $\gets$ filling$-1$
\EndIf
\EndIf
    \State $i \gets i+1$
\EndWhile
\State{\textbf{Results}: operation, qubit, address}
\end{algorithmic}
\end{small}
\end{algorithm*}

\begin{figure}
  \centering
  \includegraphics[width=6cm]{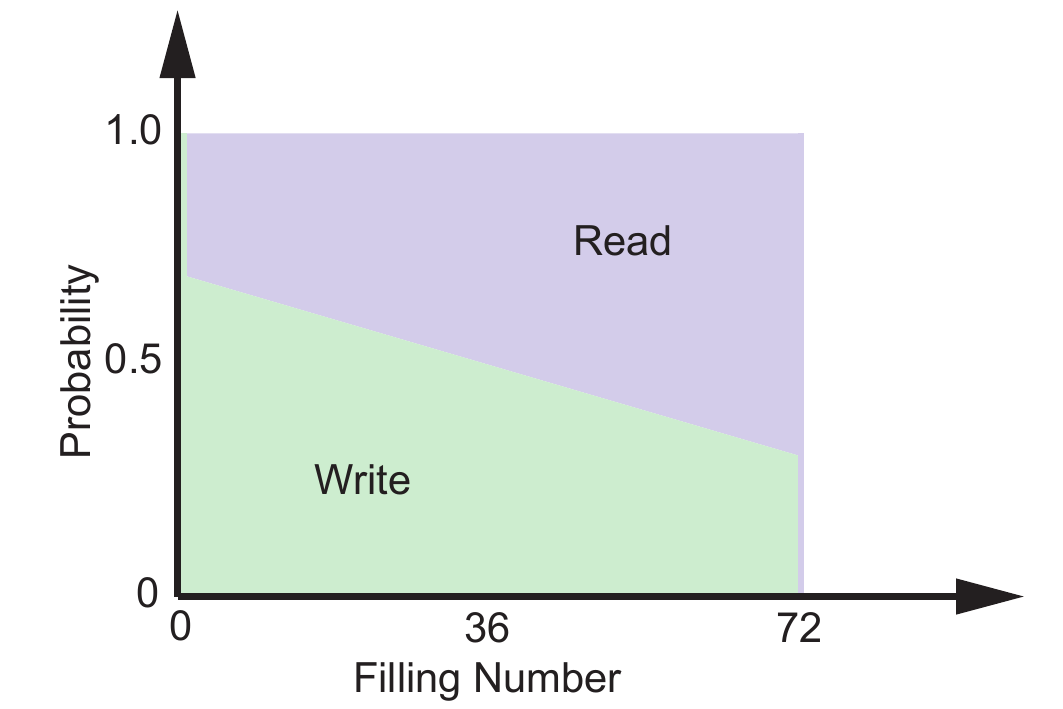}
  \caption{\textbf{The probability for determining write or read operations.} The write probability is 1 when the filling number is 0, while the read probability is 1 when the filling number is 72. In other cases, the probability is $0.65-0.3n/72$ where $n$ is the filling number.}
\end{figure}

\section{Proposed scheme to output two photons simultaneously}

The AODs can only address a single microensemble at a time, and there is only one output fiber in our current setup, thus simultaneously output of two photons is difficult. Here we can make this possible by adding a switch network to our setup in the future, as illustrated in Fig. 10. With the help of an optical switch and a delay fiber in the switch network, we can implement simultaneous output of two stored photons.

\begin{figure}
  \centering
  \includegraphics[width=8.7cm]{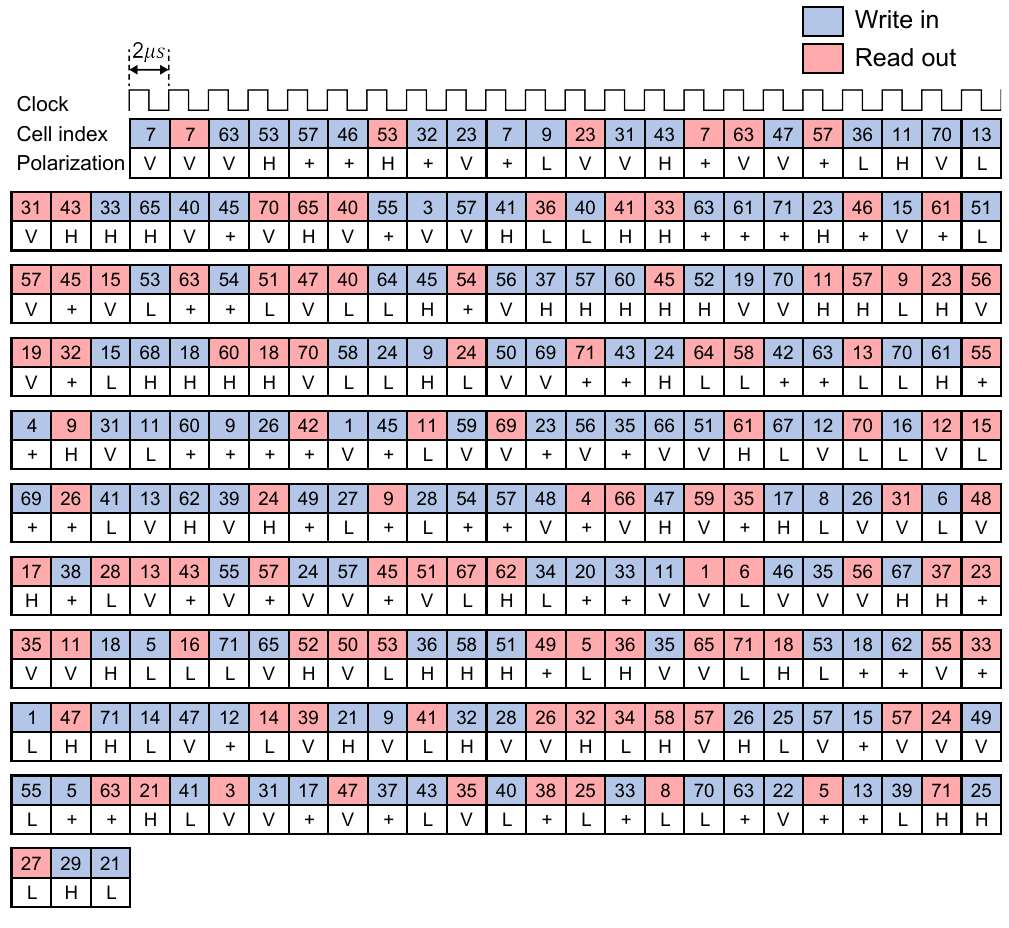}
  \caption{The sequence for RAQM with 250 operations. The detailed sequence with all the domains in Fig. 3(d) in main text.}
\end{figure}

\begin{figure}
  \centering
  \includegraphics[width=8.7cm]{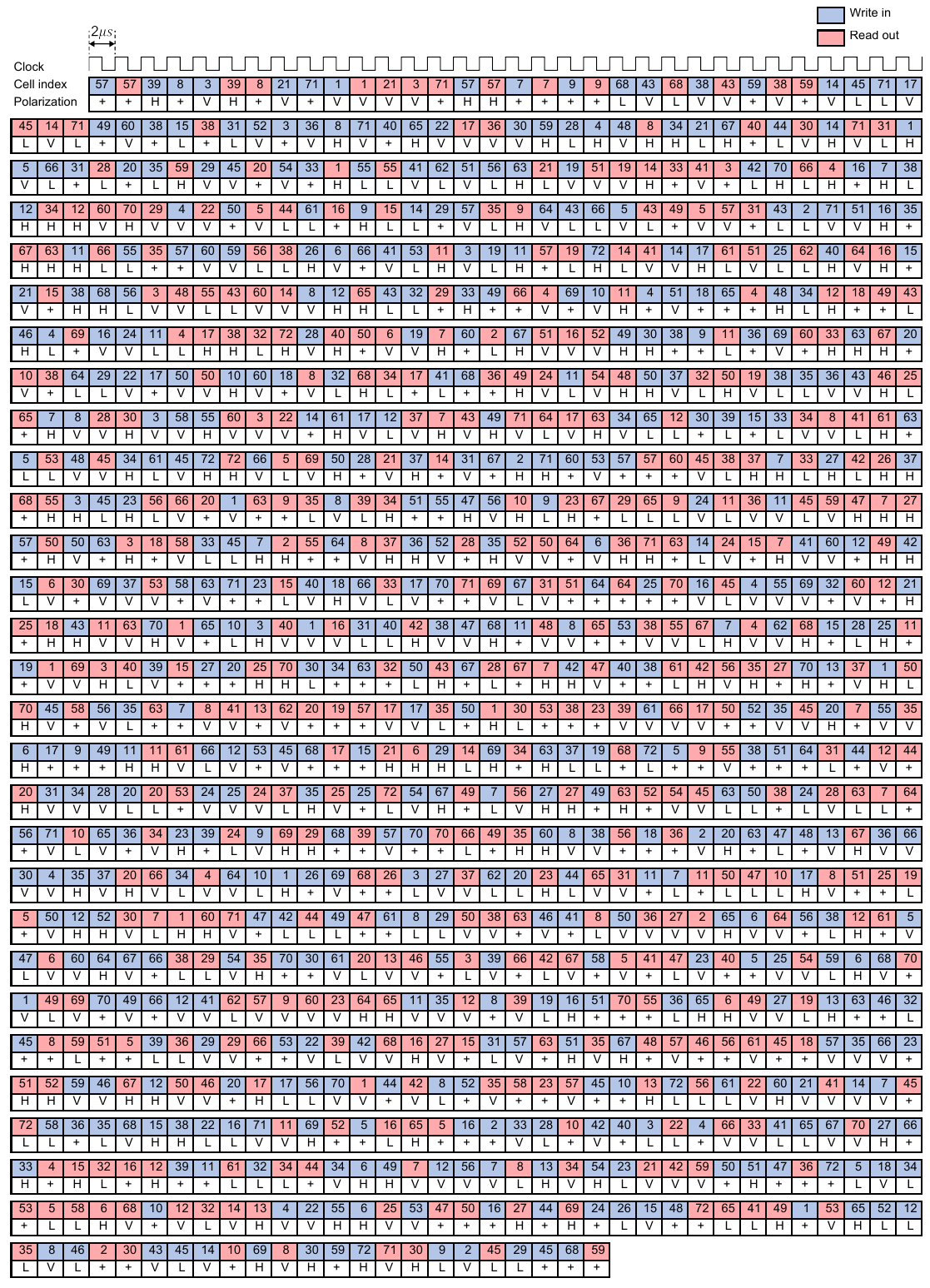}
  \caption{The sequence for RAQM with 1000 operations. The detailed sequence with all the domains in Fig. 3(e) in main text.}
\end{figure}

\section{Limiting factors for efficiency and improvement strategies}

The efficiency of the EIT storage is strongly dependent on the optical depth of the memory cell (about $5$ in the center and gradually decreases to about $3$ at the corner). This is the main reason for the efficiency inhomogeneity in different memory cells across the array. A minor source of this inhomogeneity is the degradation in the AOD mode matching efficiency (about $85\%$ in the center and $70\%$ at the corner, can be improved by better adjustments of optics) which characterizes the overlap of the input and output spatial modes. These efficiency inhomogeneities could be mitigated by more homogeneous distribution of the optical depth in each mode with a larger-sized atomic cloud, as well as better adjustments of the AOD addressing optics in the future.

To further improve the efficiency while keeping the current features, we can load the atoms into a 2D optical-lattice array to achieve high optical depth in each cell. We estimate that an EIT efficiency around $50\%$ can be reached with a moderate optical depth around $50$. A potential fundamental limitation can be induced by four-wave mixing when the optical depth gets very high (OD$>300$), which leads to noise in the output spatial mode and limits the fidelity of storage (\cite{hsiao}). However this limit can be largely suppressed via phase mismatching. EIT storage efficiencies over $85\%$ and fidelity over $99\%$ have been achieved at the same time in \cite{zhushiliang}.

Another important factor which equally influence the efficiency in every memory cell lies in the temporal wavepacket shape of the input and coupling pulses. In our experiment, the input and coupling pulses are too short ($\sim200\,$ns) which induces a large frequency broadening ($\sim5\,$MHz) comparable to the EIT transparency window, and this reduces the storage efficiency. Besides, wavepacket shaping is not applied in our experiment as the input and coupling light pulses are both square pulses, meanwhile a Gaussian shaped pulse is found to be necessary for the optimal efficiency (see \cite{zhushiliang}). This limiting factor can be solved in the future with careful wavepacket shaping and longer pulse length.

We also add a detailed list of the efficiencies of all the components in the whole optical path in Table I. The efficiency can be improved in the future with better transmission in all the optics (by using anti-reflection coated fiber, better transmission in fiber beamsplitter thus better transmission in Encoding converter, etc.)

\section{Control sequences for the RAQM}

To characterize the performance of the RAQM, we generate a $500\,\mu$s sequence including $250$ random access operations. The algorithm for generating each sequence is summarized in the pseudocode in Algorithm~$1$. The RAQM runs with a clock cycle of $2\,\mu$s, which is also the access time for one qubit cell. At every rising edge of the memory clock, we first choose whether to write a qubit into the memory or to read out a qubit from the memory. The prerequisite for this choice is that the memory array is neither empty nor full. The probability of write-in or read-out is dependent on the filling rate of the array, as shown in Fig.~11. The consideration for this probability distribution is that we want to write in qubits as quickly as possible when the array is relatively empty, but read out qubits when the array is nearly saturated. This causes the filling number to oscillate around the half, as shown in Fig. 3(d) and 3(e) of the main text. It should be noted that this probability distribution is not a necessary condition and all self-consistent sequences can run on our RAQM. If the operation is to write in a qubit, further choices are to select one qubit cell from currently unfilled cells and the polarization of the input qubit from $\{\ket{H},\ket{V},\ket{+},\ket{L}\}$, with equal chances. If the operation is to read out a qubit, we just need to choose the address of the qubit cell from the occupied cells with equal chances.

In the further demonstration of the $2\,$ms sequences with $1000$ random access operations, we maintain a $500\,\mu$s scrolling window to ensure each stored qubit stays within the coherence. We monitor if the storage time of any stored qubit is about to exceed $500\,\mu$s. If the answer is yes, the next operation is to read the qubit out. The detailed sequences for these two cases used in the main text are illustrated in Fig.~12 and Fig.~13. Due to the free falling and thermal expansion of the atomic cloud in the experiment, the memory array can survive as long as $5\,$ms. Thus we set the maximum operation number to 1000, which corresponds to $2\,$ms of experiment. This time can be elongated by setting the optical paths along the gravity direction, as well as better cooling of the atomic cloud. In the future, we expect to greatly improve the operation capacity by trapping micro-ensembles in optical lattice arrays \cite{ensemble array}. Moreover, repeated write or read operations on the same cell do not affect its later performance including retrieval efficiency and storage fidelity.

\begin{figure}
  \centering
  \includegraphics[width=8.7cm]{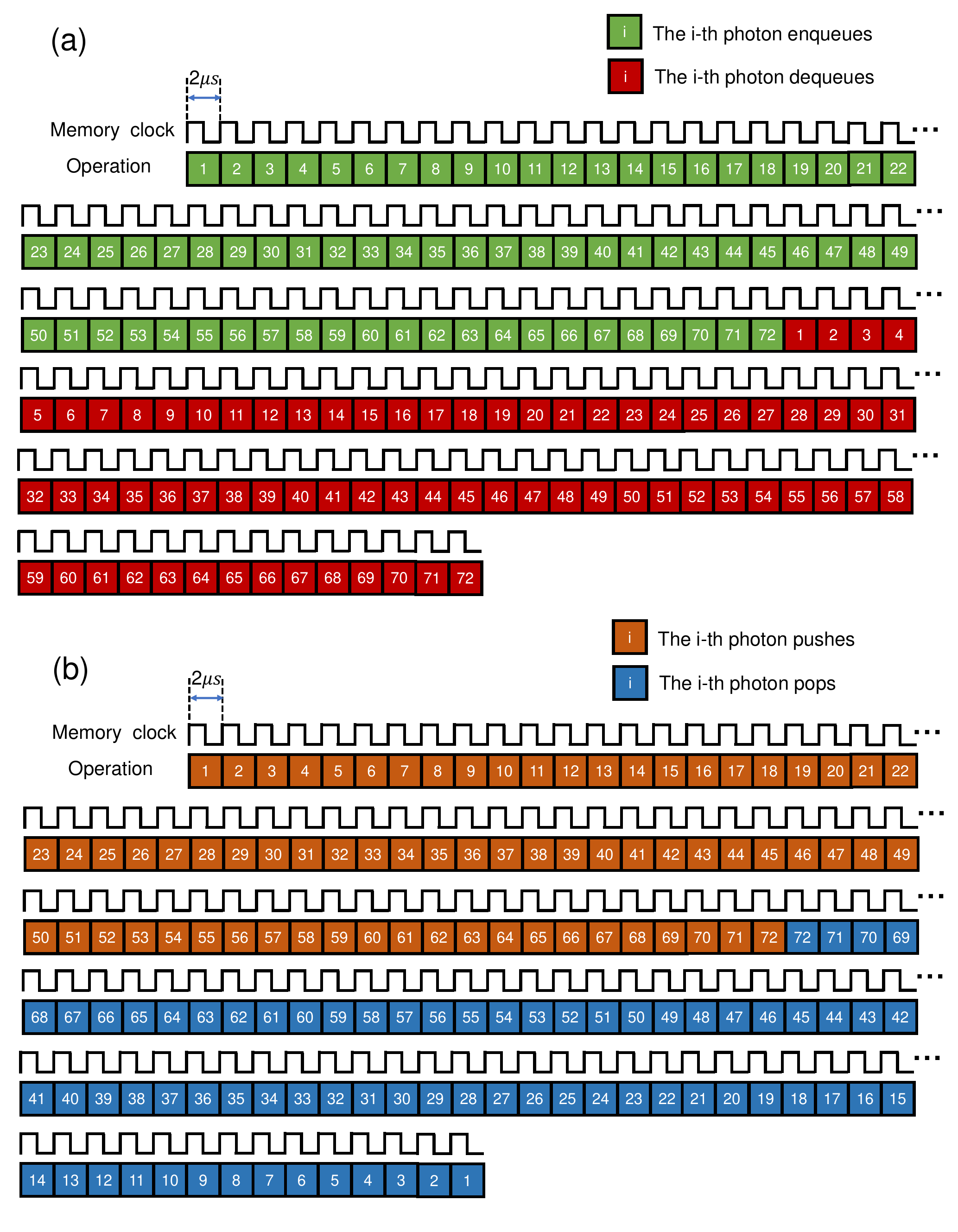}
  \caption{The sequence for quantum queue and stack in the main text. (a) The sequence for quantum queue in the main text (Fig. 4(b)-(e)). (b) The sequence for quantum stack in the main text (Fig. 4(g)-(j)).}
\end{figure}

\begin{figure}
  \centering
  \includegraphics[width=8.7cm]{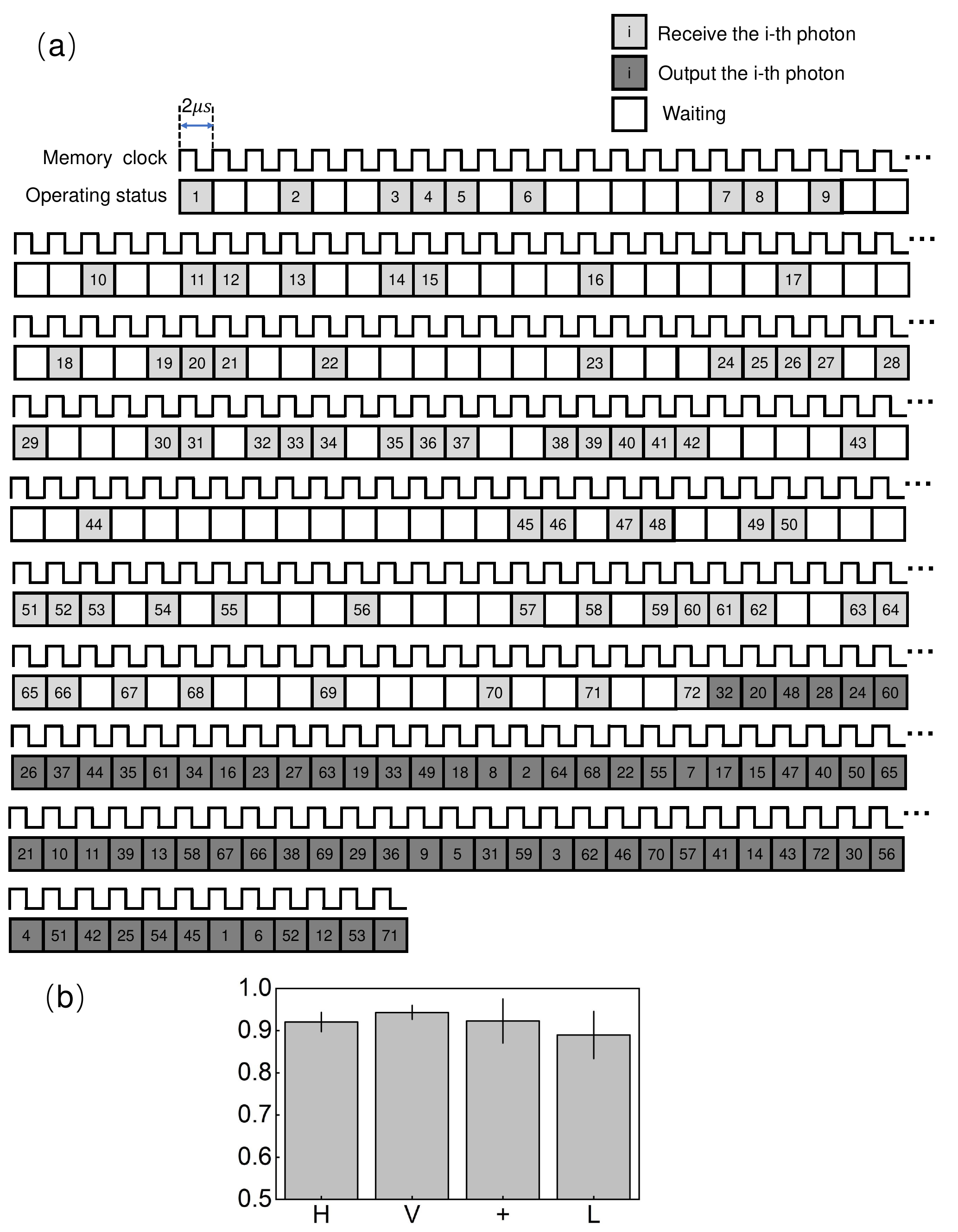}
  \caption{The sequence and fidelity for quantum buffer in the main text. (a) The sequence for quantum buffer in the main text (Fig. 4(l)-(o)). (b) The fidelities for each input polarization. The error bars represent one standard deviation.}
\end{figure}

\section{Quantum queue, stack, and buffer}

Here we present the detailed sequences for the quantum queue, stack, and buffer used in the main text, as shown in Fig.~14 and 15. In the realization of quantum queue, stack, and buffer, the $i$-th input photonic qubit is stored in the $i$-th qubit cell. The memory can simultaneously store 72 photonic qubits at maximum.

In the main text, we show two special cases of quantum queue and stack, in which the $72$ flying qubits fill the memory array at first, and then are read out. As shown in Fig. 16 and 17, we demonstrate the quantum queue and stack containing $72$ flying qubits with a general sequence in which the inputs and outputs are interweaved. At every rising edge of the memory clock ($500\,$kHz), we decide whether to input a qubit or output a qubit with equal chances. As in the main text, here we present the input distribution over $4$ polarizations, the real-time filling number of the memory array, the storage time of each photon, and the fidelity of each retrieved photon.

\begin{figure}
  \centering
  \includegraphics[width=8.7cm]{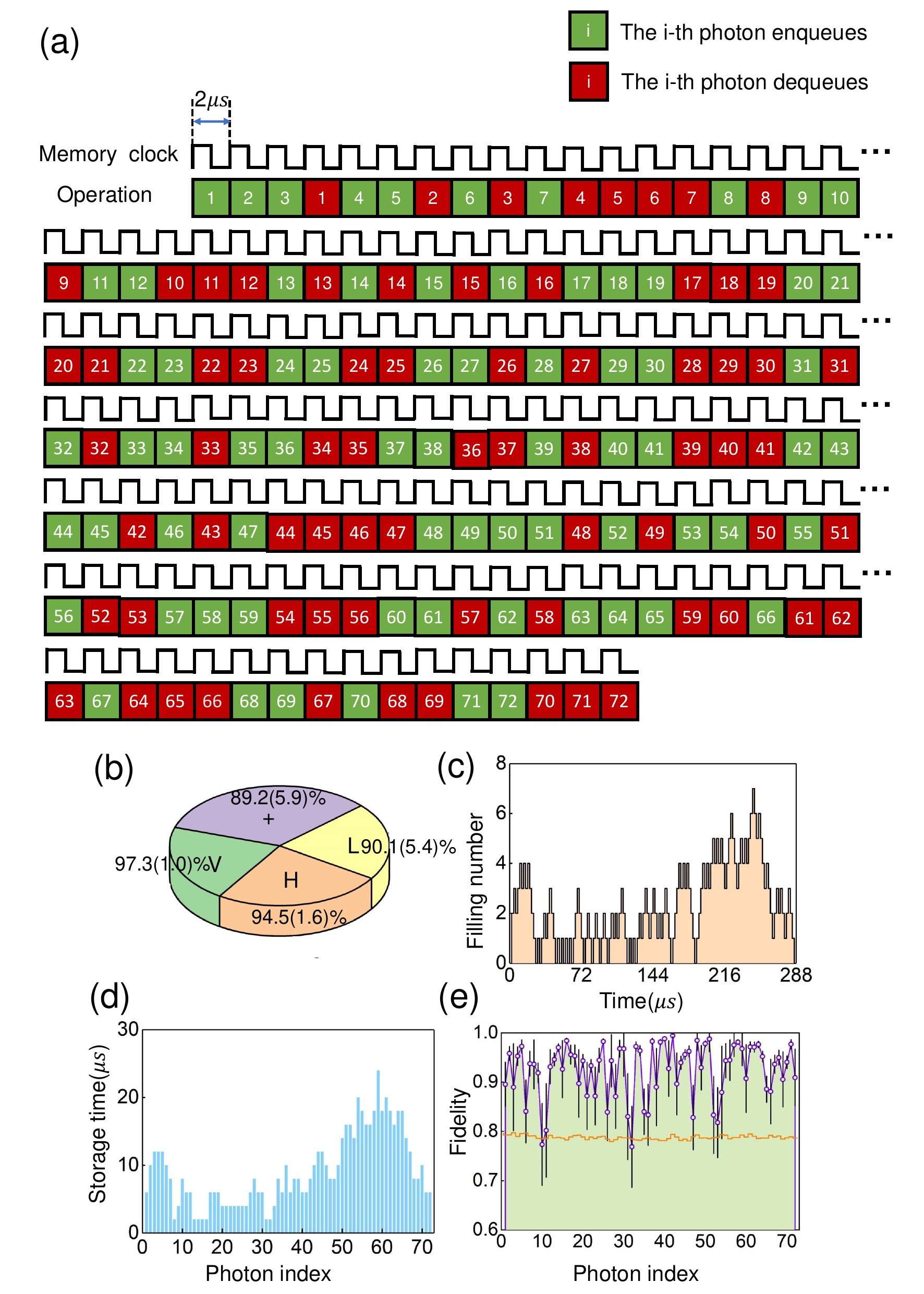}
  \caption{A general sequence for quantum queue. (a) A randomly generated sequence for quantum queue. (b) The input distribution and fidelity of  $4$ polarizations. (c) The filling number of the memory array during the experiment. (d) The storage time of $72$ photons. (e) The fidelities of the retrieved photons and corresponding quantum bound (the orange lines). The error bars represent one standard deviation in both (b) and (e).}
\end{figure}

\begin{figure}
  \centering
  \includegraphics[width=8.7cm]{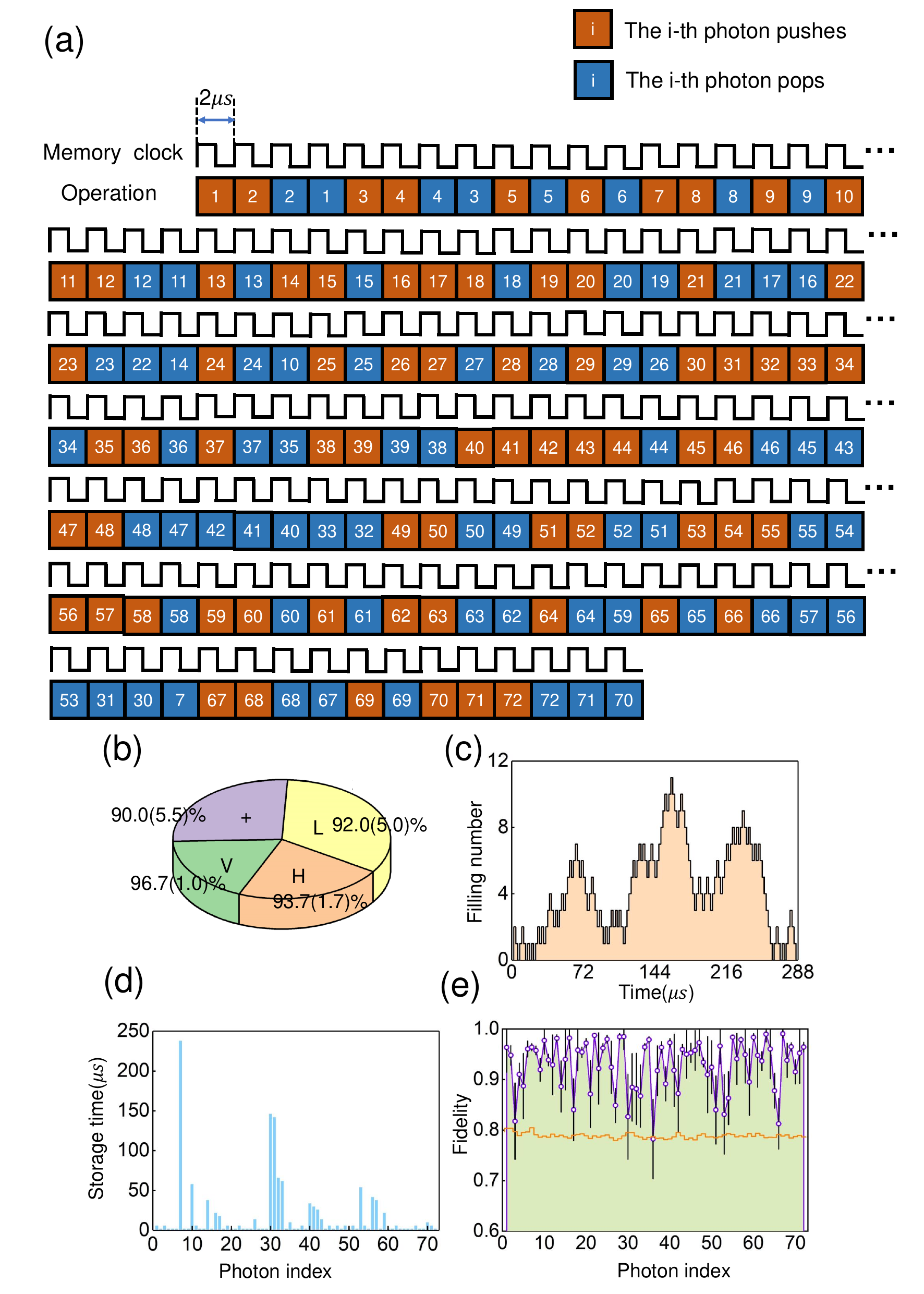}
  \caption{A general sequence for quantum stack. (a) A randomly generated sequence for quantum stack. (b) The input distribution and fidelity of  $4$ polarizations. (c) The filling number of the memory array during the experiment. (d) The storage time of $72$ photons. (e) The fidelities of the retrieved photons and corresponding quantum bound (the orange lines). The error bars represent one standard deviation in both (b) and (e).}
\end{figure}

\section{DLCZ Entanglement Source}

In this section, we describe the details of the entanglement source where a variation of the DLCZ protocol is used to generate the entangled photon pairs encoded in polarization \cite{lattice}. The setup of the entanglement source is shown in Fig.~18(a). The atomic ensemble is initially prepared to $|5S_{1/2},F=1\rangle$ state by optical pumping. A $100\,$ns weak write pulse drives the $|5S_{1/2},F=1\rangle\rightarrow|5P_{1/2},F=2\rangle$ transition and is focused on the atomic ensemble with a beam waist of $230\,\mu$m. The frequency detuning between the write laser pulse and the transition is $20\,$MHz. As shown in Fig.~18(a), we collect the scattered signal photon along two symmetric spatial directions and combine the two possible signal paths with a polarization beam splitter (PBS). The excitation probability is $1.1\%$. Once a signal photon is detected, the atom-photon entangled state can be written as
\begin{equation}
  |\Psi\rangle_{at-ph}=\frac{1}{\sqrt{2}} (\ket{L}\ket{H}+e^{i\phi_s}\ket{R}\ket{V})
\end{equation}
where $\ket{L/R}$ represents the spin wave excitation in the $L/R$ spatial mode, $\ket{H/V}$ represents the signal photon with horizontal/vertical polarization, $\phi_s$ is the relative phase caused by the path difference between two signal paths. After a controllable delay, a strong read laser couples the $\ket{5S_{1/2},F=2}\rightarrow\ket{5P_{1/2},F=2}$ transition and coherently converts the spin wave excitation to an idler photon. The entanglement between signal and idler photon is encoded in polarization and can be written as
\begin{equation}
  \ket{\Psi}_{s-i}=\frac{1}{\sqrt{2}}(\ket{H}\ket{V}+e^{i(\phi_s+\phi_i)}\ket{V}\ket{H})
\end{equation}
where the relative phase $(\phi_s+\phi_i)$ is compensated to be zero in experiment. The diamond optical configuration constitutes a MZ interferometer and the path difference is actively stabilized by real-time feedback. The lock laser is turned off during the entanglement generation to avoid destroying the atomic internal state and to eliminate the noise. We reduce the write-clean cycle to $700\,$ns by compressing all delays in the system, as shown in Fig. 18(b). With the excitation probability set to 1.1\%, we can generate an entangled signal-idler photon pair from the source in about $70\,\mu$s, and 4 pairs of entangled signal-idler photons in $500\,\mu$s with $93.5\%$ probability. We estimate the entanglement fidelity of the signal-idler photon pairs from the source to be $94(1)\%$ through quantum state tomography. The reconstructed density matrix for the signal-idler entanglement is shown in Fig.~18(c). The real-time control is implemented by a $100\,$MHz home-made field-programming-gate-array (FPGA) board.

\section{Storage and reshuffle of $4$ heralded quantum entanglements}

\begin{figure}
  \centering
  \includegraphics[width=8.7cm]{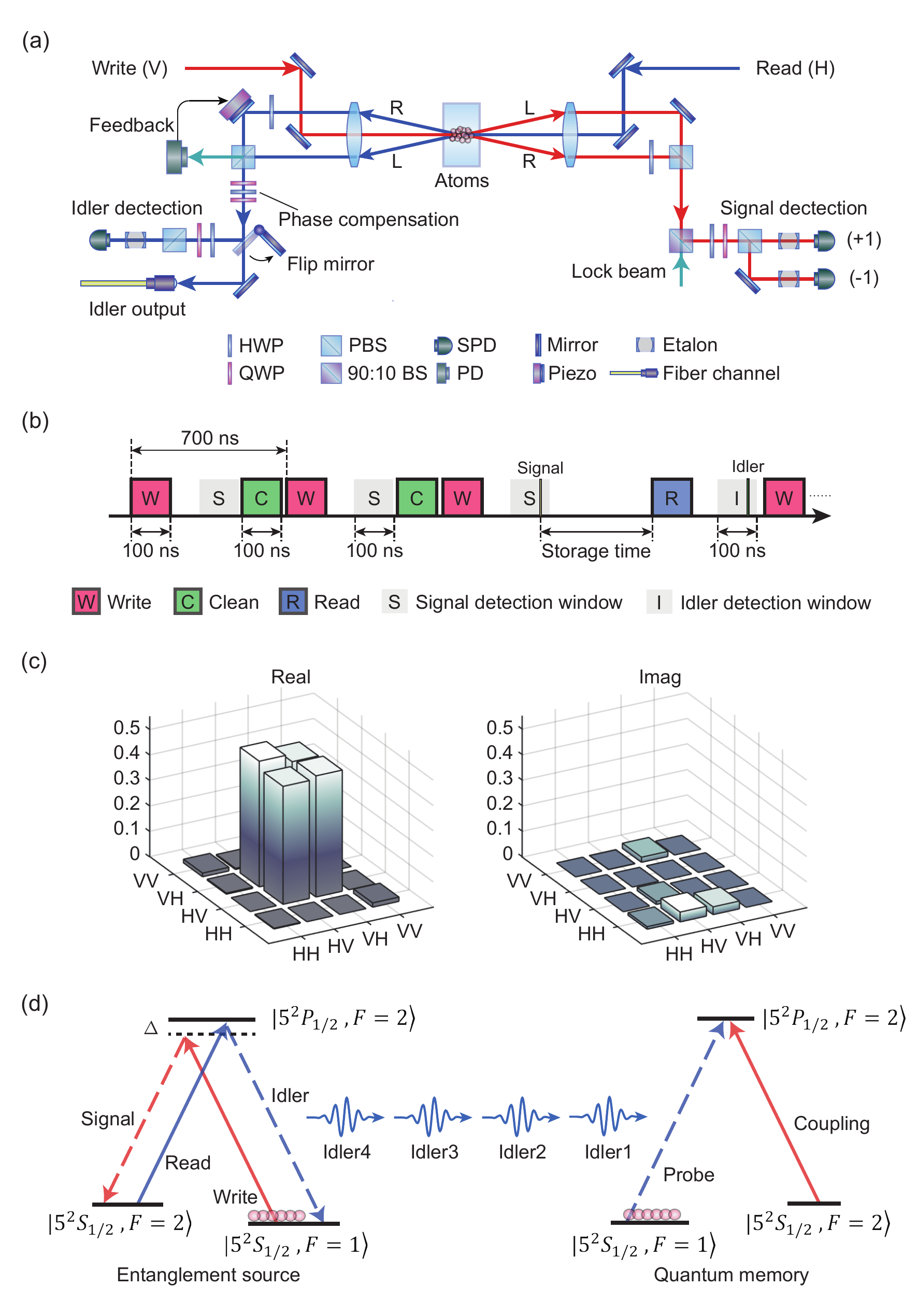}
  \caption{The entanglement source and energy levels. (a) The experimental setup. (b) The experimental sequence for the entanglement generation. (c) The reconstructed density matrix for the signal-idler entangled state. (d) The energy level diagram for entanglement generation, catch, freezing, and release. }
\end{figure}

Here we describe the detailed sequence for the storage and reshuffle of $4$ heralded entanglements. As shown in Fig.~19, we start with repeated write-clean cycles in the entanglement source based on the DLCZ protocol. Once a signal photon (signal $1$) is recorded, we immediately convert the spin wave excitation into an idler photon (idler $1$) and store idler $1$ in the multipurpose quantum memory. We repeat this process $4$ times and store $4$ idler photons into the multipurpose quantum memory. Note that due to the probabilistic nature of the entangled photon source, all the $4$ idler photons are captured at random times. After the  catch and storage of $4$ entanglements, we retrieve the $4$ stored qubits out with any desired order to demonstrate the reshuffle. Here we use the order $2$-$4$-$1$-$3$ as an example, which means that the $2$nd captured idler photon (idler $2$) will be first read out from the memory. The real-time feed-forward process is implemented by an FPGA which dynamically select one of the desired waveforms from AWGs conditioned on the detection events.

We then evaluate the fidelities of the $4$ reshuffled entanglements \cite{5}. An arbitrary two-qubit density matrix can be expressed as
\begin{equation}
  \rho=\frac{1}{4}\sum_{i,j} \rho_{i,j} \sigma_i\otimes\sigma_j
\end{equation}

where $i,j=\{0,x,y,z\}$ and $\sigma_i$ represents the corresponding Pauli matrix.
In our experiment, the target state is a Bell state $\ket{\Psi^+}=\frac{1}{\sqrt{2}}(\ket{H}\ket{V}+\ket{V}\ket{H})=\frac{1}{\sqrt{2}}(0,1,1,0)^T$,
so the estimated fidelity is
\begin{equation}
  F=\frac{1+\rho_{xx}+\rho_{yy}-\rho_{zz}}{4}.
\end{equation}

The coefficient $\rho_{ij}=tr(\rho\sigma_i\otimes\sigma_j)$ can be obtained through projective measurements under Pauli bases. In our experiment, both signal and idler photons are projected by different measurement bases and detected by a pair of SPDs, which represents $+1$ and $-1$, respectively. We collect the coincidence counts under $3$ measurement bases $\{\sigma_x\otimes\sigma_x,\sigma_y\otimes\sigma_y,\sigma_z\otimes_z\}$, with
\begin{equation}
  \rho_{ii}=\frac{n_{++}+n_{--}-n_{+-}-n_{-+}}{n_{++}+n_{--}+n_{+-}+n_{-+}}
\end{equation}
where $n_{++}$, $n_{+-}$, $n_{-+}$, $n_{--}$ are the coincidence counts. The fidelities of 4 reshuffled entangled photon pairs are eventually estimated to be 83.5\% (S1-I1), 87.7\% (S2-I2), 91.9\% (S3-I3), and 89.0\% (S4-I4).

\begin{figure}
  \centering
  \includegraphics[width=8.7cm]{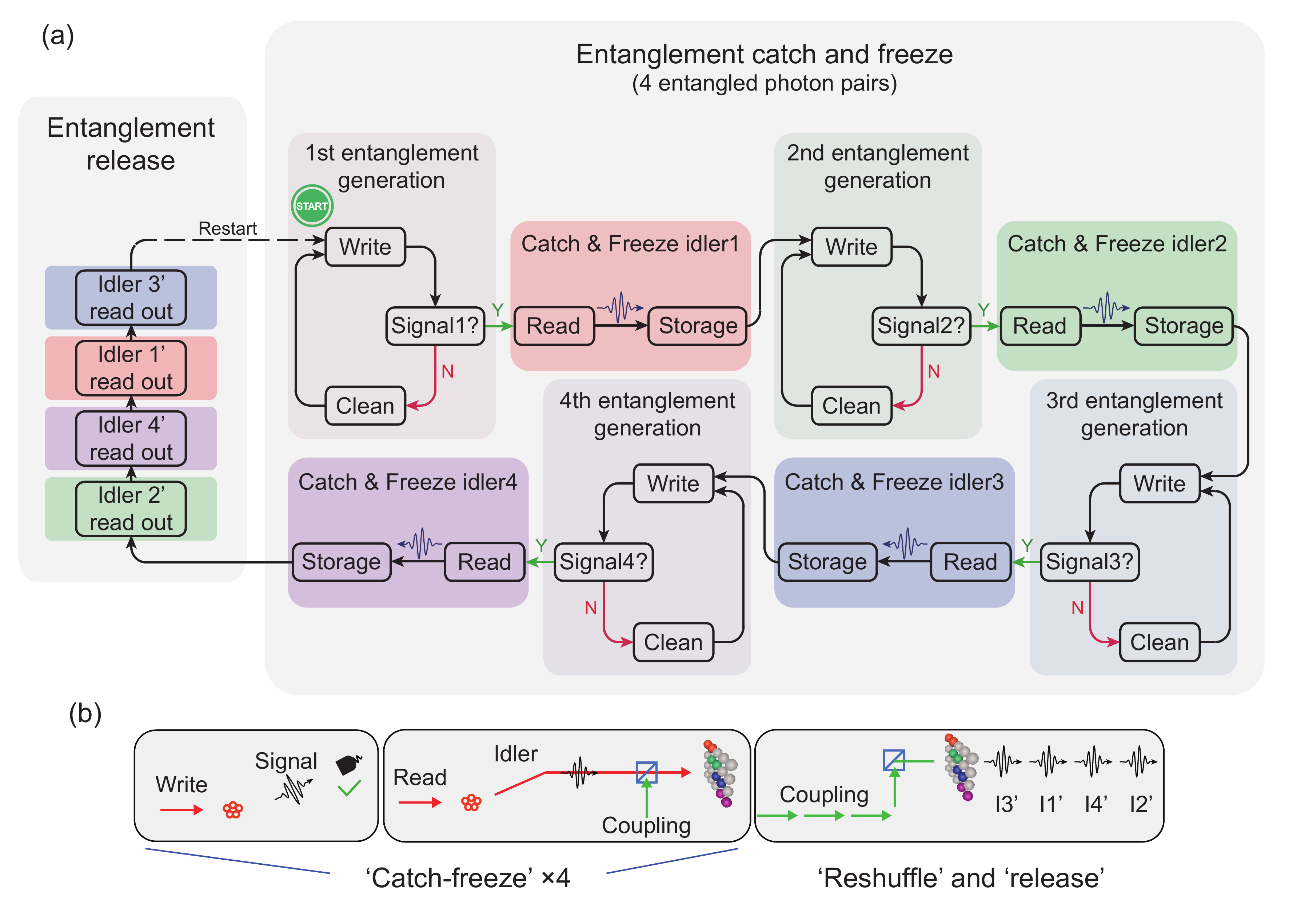}
  \caption{The sequence for the storage and reshuffle of 4 heralded quantum entanglements. (a) The detailed sequence of the experiment. The real-time feed-forward is implemented on an FPGA. (b) A summarised sequence.}
\end{figure}

%
%
%
%
%

\end{document}